\documentclass[aps,pre,amsmath,amssymb,floatfix,preprint,nofootinbib]{revtex4}
\usepackage{graphicx}
\usepackage{amsmath}

\begin{document}
\title{Derivation of Markov processes that violate detailed balance}
\author{Julian Lee}
\email{jul@ssu.ac.kr}
\affiliation{Department of Bioinformatics and Life Science, Soongsil University, Seoul, Korea}
\date{\today}
\begin{abstract}
Time-reversal symmetry of the microscopic laws dictates that the equilibrium distribution of a stochastic process must obey the condition of detailed balance. However, cyclic Markov processes that do not admit equilibrium distributions with detailed balance are often used to model systems driven out of equilibrium by external agents. I show that for a Markov model without detailed balance, an extended Markov model can be constructed, which explicitly includes the degrees of freedom for the driving agent and satisfies the detailed balance condition. The original cyclic Markov model for the driven system is then recovered as an approximation at early times by summing over the degrees of freedom for the driving agent. I also show that the widely accepted expression for the entropy production in a cyclic Markov model is actually a time derivative of an entropy component in the extended model. Further, I present an analytic expression for the entropy component that is hidden in the cyclic Markov model. 
\end{abstract}
\maketitle
\section{Introduction}
A Markov process\cite{kamp,chung} is a paradigmatic model for describing a stochastic process in various fields of science including biophysics\cite{ken07,ken15,gop,sung2,brown,feng,Pande1,chodera,Chu,Hyeon}. A Markov process can be obtained from the microscopic dynamics of a closed system by coarse-graining\cite{groot,zwan}. Thus, it can be considered as a process where information is continuously being lost. Alternatively, Markov processes have also been obtained by maximizing the dynamical entropy of the probability distribution of stochastic paths under appropriate constraints\cite{rmp,ken,sl}.

Considering discrete states labelled by an index $i$, the time evolution of a probability distribution in a Markov jump process, where transitions occur only at times that are integer multiples of $\Delta t$, is given by
~\cite{markov1}
\begin{equation}
\pi_i\left(t + \Delta t \right) = \sum_j \pi_j(t) p_{j \to i},  \label{markov2}
\end{equation}
where  $\pi_i(t)$ is the probability that the system is in a state $i$ at time $t=n\Delta t$ for some integer $n$, and $p_{i \to j}$ is the transition probability from the state $i$ to $j$. The conservation of probability implies that $\sum_j p_{i \to j} = 1$. Eq.(\ref{markov2}) can also be written as
\begin{equation}
\frac{\Delta \pi_i(t)}{\Delta t} = \sum_j \pi_j(t ) k_{j \to i},  \label{mard}
\end{equation}
where $\Delta \pi_i(t) \equiv \pi_i(t + \Delta t) - \pi_i(t)$ and
\begin{equation}
k_{i \to j} \equiv \frac{ p_{i \to j}-\delta_{i,j}}{\Delta t},\label{rate}
\end{equation}
with $\delta_{i,j}$ being the Kronecker delta, which is one if $i= j$ and zero otherwise. Here, $k_{i \to j}$ is called the transition rate from the state $i$ to $j$ for $i \ne j$.  The conservation of probability imposes the constraint that $\sum_j k_{i \to j} = 0$, from which we obtain $k_{i \to i} = -\sum_{j \ne i} k_{i \to j}$. The transition rates completely determine the stochastic evolution of the system. The values of $k_{i \to j}$ are considered to be time-independent constants, and the term time-homogeneous Markov process is sometimes used to emphasize this fact.

 The equation for the continuous-time Marvkov process is obtained from Eq.(\ref{mard}) by taking the limit $\Delta t \to 0$:
\begin{equation}
\frac{d \pi_i(t)}{dt} = \sum_j  \pi_j(t) k_{j \to i} ,\label{markov}
\end{equation}
which is called the master equation. 
  
 It is a well-known fact that under appropriate conditions, the probability distribution of a Markov chain converges to a unique stationary distribution $\pi^{\rm st}_i$ regardless of the initial distribution~\cite{markov1}. A stationary distribution satisfies the balance condition:
\begin{equation}
\sum_j  \pi^{\rm st}_j k_{j \to i}=\sum_j \left[ \pi^{\rm st}_j k_{j \to i} -  \pi^{\rm st}_i k_{i \to j} \right] = 0\quad \forall i , \label{balance}
\end{equation}
where the second expression follows from the first by the conservation of probability, $\sum_j k_{i \to j} =0$. 
  A stationary distribution is considered a true equilibrium, which we now denote as $\pi^{\rm eq}_i$, only if a stronger condition called detailed balance holds:
\begin{equation}
\pi^{\rm eq}_j k_{j \to i} -  \pi^{\rm eq}_i k_{i \to j} =0\quad \forall i,j. \label{db}
\end{equation}
  
   A given Markovian transition matrix admits an equilibrium solution with detailed balance if and only if Kolmogorov's criterion is satisfied. It states that for any cycle of states $i_0, i_1, \cdots, i_n, i_0$, the product of forward transition rates over the cycle is equal to that of the reverse rates~\cite{kogo,kelly}:
\begin{equation}
k_{i_0 \to i_1} k_{i_1 \to i_2} \cdots k_{i_{n-1} \to i_n} k_{i_{n} \to i_0} 
=  k_{i_0 \to i_n} k_{i_{n} \to i_{n-1} } \cdots k_{i_2 \to i_1}  k_{i_1 \to i_0} .\label{kobo}
\end{equation}
Therefore, we see that the existence of a cycle in the network topology of a Markov process is a necessary condition that its stationary distribution violates the detailed balance condition. From here on, we will call a Markov model that violates Kolmogorov's criterion simply a cyclic Markov model, since cycles that satisfy Kolmogorov's criterion are not of interest here. 

Because a Markov process is a coarse-grained description, a state labelled with the index $i$ is usually not a true microstate of  the closed system, but rather an aggregate of such microstates. 
The crucial assumption underlying the coarse-graining that leads to Eq.(\ref{markov}) is that of instantaneous local equilibrium: the equilibration between the microstates within each Markov state occurs much faster than the transition between distinct Markov states. Therefore, we may call the index labelling the Markov states as the slow variable, and the underlying additional hidden index required for specifying the microstate as the fast variable.
Once instantaneous local equilibrium is assumed, the detailed balance condition for an equilibrium distribution follows from the symmetry of the underlying microscopic laws under time reversal, under the condition that the index $i$ is invariant under time reversal, which will be assumed always true in this paper\cite{groot,zwan}. This suggests that any closed system can be described by a Markov process that satisfies Kolmogorov's criterion if coarse-graining is performed properly. However, cyclic Markov processes that violate Kolmogorov's criterion are often used to model systems continuously driven out of equilibrium by an external agent\cite{BS,cycle1,cycle2,wang,QH11,QH12,QH2,QHx1,QHx2,GQ1}. The stationary state of such a model is called the nonequilibrium steady state\cite{GQ2,QH11,QH12,QH2,QHx1,QHx2,GQ1,QH31,QH32,hi12} because the detailed balance condition does not hold. 
 
 It has been argued that these rather contradictory views can be reconciled if the cyclic Markov process is embedded in a larger Markov model that explicitly includes the degrees of freedom for the driving agent\cite{zmh}. Obviously, the total system consisting of the driven system plus the outside environment containing the driving agent forms  a closed system, which will eventually reach equilibrium. For example, a cyclic Markov model can be used to describe a biochemical cycle driven by ATP. However, from a more global point of view, the cycle will stop once all ATP molecules are used up. If we consider a situation where ATP itself is regenerated by food intake, we know that the cycle is still a part of a larger cycle driven by the sun. Considering a closed system that includes all biological organisms as well as the sun, the whole system will reach equilibrium once the sun has burnt out and all life processes have ended. Therefore, the dynamics for the driven system are described by a model where the transition rate is time-dependent. Since the transition rates change with time, it is possible that the rates violate Kolmogorov's criterion at earlier times, but satisfy the criterion as the system reaches equilibrium. A cyclic Markov process is clearly only an approximate description valid only for time periods much earlier than the equilibration of the total closed system. 
 
 In this work, it is shown that for any time-homogeneous Markov model without detailed balance, an extended Markov model that explicitly includes the degrees of freedom for the driving agent and satisfies Kolmogorov's criterion can be constructed. The original cyclic Markov model for the driven system is then recovered as an approximation at early times by summing over the degrees of freedom of the driving agent. By constructing the extended model, the widely accepted formula for the entropy production in a cyclic Markov model is explicitly expressed as a time derivative of an entropy component. Furthermore, an analytic expression for the entropy component is presented, which is hidden in the original cyclic Markov model. 
 
\section{Derivation of Markov model that violates detailed balance}

\subsection{Three-state model}
 Before providing a derivation for general Markov processes without detailed balance, a simple example of a discrete-time Markov process is presented, consisting of the three states shown in Figure 1(a), where $a b c \ne \alpha \beta \gamma$. Now, consider an extended model where the state of the driving agent, labelled by an integer $X\ (0 \le X \le N)$, is explicitly included. We assume that the change of $X$ is uniquely determined for each $i \to j$ transition, denoted by $\Delta X (i \to j)$. We also assume that $\Delta X (i \to j) = -\Delta X (j \to i)$. For example, this three-state process may be a biochemical cycle driven by the hydrolysis of ATP to ADP. Then, we may take $N$ to be the total number of ATP and ADP molecules, which is assumed to be fixed, and let $X$ and $N-X$ be the numbers of ATP and ADP molecules, respectively. In this case, $-\Delta X (i \to j)$ is the number of consumed ATP molecules in the biochemical reaction $i \to j$. We request that the sum of $\Delta X(i \to j)$ along the cycle is nonzero. This leads to an absence of any cycle in the extended model, which in turn guarantees the detailed balance. There is no unique extended model corresponding to the cyclic Markov model in Figure 1(a). For example, we may have $\Delta X = \pm 1$ for each transition (Figure 1(b)), or alternatively  $\Delta X = \pm 1$  only for the transitions between C and A and $\Delta X = 0$ for all the other transitions (Figure 1(c)). However, the model in Figure 1(c) can be clearly mapped into that of Figure 1(b) by redefining the coordinate $X$ and changing the value of $N$. Hence, these models are mathematically equivalent. From here on, $X$ will be defined as in Figure 1(b), so that there are a total of $N+1$ states in the extended model. Then, $X$ is just a serial number attached to the states in the extended model, and does not necessarily coincide with the number of ATP molecules.

We now consider a Markov process for the probability distribution $\Pi_{(i,X)}(t)$ of the extended system, and assume that the transition probability $P_{(i,X) \to (j,Y)}$ from $(i,X)$ to $(j,Y)$ in the extended model has the form
\begin{equation}
P_{(i,X) \to (j,Y)} = p_{i \to j}\delta(Y-X,\Delta X(i \to j)),  \label{simp}
\end{equation}
where $\delta(x,y) \equiv \delta_{x,y}$ is the Kronecker delta function. That is, the nonzero value of the transition probability depends only on $i$ and $j$ (Figure 1(b), (c)). Starting from the extended model, we now sum over the degrees of freedom $X$ to obtain the reduced model describing the time evolution of $\pi_i(t)$. The dynamics of the reduced model is not described by a time-homogeneous Markov process in general, but the transition probability $q(i \to j; t)$ can still be defined, and is given by~(Appendix \ref{trcnd})
\begin{eqnarray}
&&q(i \to j; t) \equiv \frac{{\rm Pr}(i,t; j,t+\Delta t)}{\pi_i(t)}  \nonumber\\
&=& \frac{ \sum_{X,Y}{\rm Pr}(i,X,t; j,Y,t + \Delta t)}{\pi_i(t)}\nonumber\\
&=& \frac{1}{\pi_i(t)}\sum_{X,Y} \Pi_{(i,X)}(t) P_{(i,X) \to (j,Y)}
\nonumber\\
&=& p_{i \to j} \frac{\sum_{X} \Pi_{(i,X)}(t)}{\pi_i(t)} \label{wij}
\end{eqnarray}
where ${\rm Pr}(i,t; j,t+\Delta t)$ is the joint probability that the state of the driven system is $i$ at time $t$ and $j$ at time $t+ \Delta t$. Similarly, ${\rm Pr}(i,X,t; j,Y,t+\Delta t)$ is the joint probability that the state of the extended model is $(i,X)$ at time $t$ and $(j,Y)$ at time $t+\Delta t$.
In Eq.(\ref{wij}), the first line follows from the definition of the  transition probability (Appendix \ref{trcnd}), and the same definition was used for the extended model to obtain the third line. Finally, the condition Eq.(\ref{simp}) was used to derive the last line.

 Note that $X=N$ and $X=0$ are excluded from the summation in the numerator of Eq.(\ref{wij}) for $\Delta X(i \to j) = 1$ and $\Delta X(i \to j)=-1$, respectively, leading to
\begin{equation}
	q (i \to j; t) = \left\{
	\begin{array}{ll}
		p_{i \to j}(1-\frac{\Pi_{(i,N)}(t)}{\pi_i(t)}) 	&	\quad  \Delta X(i \to j) = 1,\\
   p_{i \to j}(1-\frac{\Pi_{(i,0)}(t)}{\pi_i(t)}) 	&	\quad  \Delta X(i \to j) = -1,\\		
		p_{i \to j}	&	\quad \mathrm{otherwise}.
	\end{array}
	\right. \label{reduced}
\end{equation}
As mentioned earlier, the coarse-graining of microscopic dynamics under appropriate conditions leads to a time-homogeneous Markov model that satisfies Kolmogorov's criterion. From Eq.(\ref{reduced}), we now see why the cyclic Markov model for the three states violates Kolmogorov's criterion:  We cannot assume instantaneous equilibration among the microstates within a state labelled by the index $i$, because the dynamics of the variable $X$ is not fast enough. Only in the limit of $t \to \infty$, $\Pi_{(i,X)}(t)$ approaches the equilibrium distribution where the time-dependence in Eq.(\ref{reduced}) disappears, leading to a time-homogeneous Markov model that satisfies Kolmogorov's criterion. It is straightforward to obtain the analytic form for the equilibrium solution by using the detailed-balance condition~(Appendix \ref{eqsol}).

Now consider the early time period. It is clear that if $N \gg 1$ and $\Pi_{(i,X)}(0)$ are nonzero only around the intermediate values of $X$, say $N/2$, then both $\Pi_{(i,0)}(t)$ and $\Pi_{(i,N)}(t)$ remain negligible for $t \ll t_{\rm eq}$, where $t_{\rm eq} = N \Delta t/p$ is the time scale of equilibration, with $p$ denoting the typical size of $p_{i \to j}$. In this regime, $q(i \to j; t) \simeq p_{i \to j}$, and the time-homogeneous cyclic Markov model with broken detailed balance is recovered. The driven system reaches the steady state of the cyclic model around $t_{\rm st} = \Delta t/p$, which is actually a quasi-steady state that persists for $t_{\rm st} \ll t \ll t_{\rm eq}$. 
 
 Let us refer to the three-state model with transition probability given by Eq.(\ref{reduced}) as model 1~(Figure 1). The result of a numerical computation for model 1 is shown in Figure 2, with a transition probability given by $p_{C \to A} = 0.5$ and $p_{i \to j} = 0.25$ for all other pairs with $i \ne j$. With $N=3000$, the system is initially in the state $(i,X)=(B,1500)$, with $X$ defined as in Figure 1(b). Note that for $t < 1500 \Delta t$, $\Pi_{i,N}(t) = \Pi_{i,0} (t)=0$, so the system is exactly described by the time-homogeneous three-state Markov model with broken detailed balance (Figure 1(a)). The steady-state distribution and the currents of the cyclic model are $(\pi^{\rm st}_A ,\pi^{\rm st}_B,\pi^{\rm st}_C) = (5/12,1/3,1/4)$, and $J^{\rm st} = 1/48$, respectively, which are actually the quasi-steady state distribution and currents of the extended model. We see that the system reaches the quasi-steady state at around $t/\Delta t \sim 4$~(Figure 2(a)). As we look at a longer time scale, we see that the system makes a transition from the nonequilibrium quasi-steady state to true equilibrium with $(\pi^{\rm eq}_A ,\pi^{\rm eq}_B,\pi^{\rm eq}_C) = (2/5,2/5,1/5)$ and $J_{\rm eq}=0$ around $t/\Delta t  \sim 25000 $~(Figure 2(b)). The three-state system is now described by a time-homogeneous Markov model with detailed balance, where $p_{C \to A} = 0.5$, $p_{B \to C} = 0.125$, and $p_{i \to j} = 0.25$ for all other pairs with $i \ne j$.
 
The condition that the nonzero values of the transition rates depend only on the states of the driven system, Eq.(\ref{simp}), may be overly strict to be realistic. We now consider a more general situation, where the nonzero values of the transition rate $P_{(i,X) \to (j,Y)}$ also depend on $X$, so that the constant $p_{i \to j}$ in Eq.(\ref{simp}) is now replaced by $p_{i \to j} (X) $, which is a function of $X$. Even in this more general case, the previous arguments presented under the condition Eq.(\ref{simp}) remain valid,  as long as $p_{i \to j}(X)$ is a slowly varying function of $X$ so that 
\begin{eqnarray}
P_{(i,X) \to (j,Y)} &=& p_{i \to j}(X)\delta(Y-X,\Delta X(i \to j))\nonumber\\
&\simeq& p_{i \to j}(X_0)\delta(Y-X,\Delta X(i \to j))  
\end{eqnarray}
for $X \ll N$. We then get $q(i \to j; t) \simeq p_{i \to j}(X_0)$ for $t \ll t_1 = N \Delta t/p(X_0)$. However, in contrast to the model where the values of $p_{i \to j}$ are constants that are independent of $X$, the system does not reach equilibrium at $t \sim t_1$, because $p_{i \to j}(X)$ deviates significantly from $p_{i \to j}(X_0)$ as $t \to t_1$. This means that $t_{\rm eq}$ is less well defined in the model with $X$-dependent values of non-zero transition probability, suggesting that the transition to equilibrium is smoother.

As a simple example, let us consider a three-state model that is more realistic than model 1, which we call model 2, where the transitions   $C \to A$ and $A \to C$ are driven by the reactions ${\rm ATP} \to {\rm ADP}+{\rm P}$ and ${\rm ADP}+{\rm P} \to {\rm ATP}$, respectively. We take $N=3000$, as in the case of model 1, where $N+1$ is the total number of states in the extended model, labelled by the coordinates $X=0, \cdots N$. As in the case of model 1, we assume that the state of the driven system at both ends of the Markov chain is B. It is then easy to see that the numbers of ATP and ADP molecules are $N_{\rm ATP}= [(X+1)/3]$ and $N_{\rm ADP} = N/3-[(X+1)/3]$, respectively, with their total number fixed as $N_{\rm tot}=N/3 =1000$, where $[X]$ denotes the integer part of the number $X$.  The result of the numerical computation for model 2 is shown in Figure 2(a) and (c), where the initial condition and parameters are the same as in the case of model 1, except that 
 $p(C \to A) = 0.50 N_{\rm ATP} /N_{\rm tot}$ and $p(A \to C) = 0.25 N_{\rm ADP}/N_{\rm tot} $ so that they are proportional to the numbers of ATP and ADP, respectively. The parameters are chosen so that they coincide with those of model 1 for the initial value of $X=X_0 \equiv 1500$.  The behavior of model 2 is almost identical to that of model 1 at early times, as expected (Figure 2 (a)). As in  model 1, the system reaches the quasi-steady state at $t_{\rm st}\sim \Delta t/p(X_0)\simeq 4 \Delta t$.  However, in contrast to model 1, the system does not make a sharp transition to equilibrium at a well-defined $t_{\rm eq}$, but rather makes a much smoother transition to equilibrium characterized by $(\pi^{\rm eq}_A ,\pi^{\rm eq}_B,\pi^{\rm eq}_C) = (1/3,1/3,1/3)$ and $J_{\rm eq}=0$, as predicted. Further details regarding the equilibrium distribution for both model 1 and model 2 can be found in Appendix \ref{eqsol}. 
 
\subsection{General derivation}
  The discussion above can be generalized to any Markov model that violates detailed balance. Now there can be more than one cycle in the Markov network, and accordingly more than one driving agent. All the degrees of freedom for the driving agent are now grouped and expressed as a vector ${\bf X}$,  where we regard the components of ${\bf X}$ to be dimensionless without loss of generality.  For example, in a realistic biochemical cycle, ATP will not simply be exhausted, but rather replenished by another biochemical cycle. This biochemical cycle may be coupled to other chemical cycles, which are ultimately coupled to radiation energy coming from the sun.  Then, the vector ${\bf X}$ represents the state of all the degrees of freedom involved in driving the biochemical cycle of interest, including the amount of hydrogen in the sun. To encompass both discrete-time models and continuous-time models, I will describe a model in terms of transition rate rather than transition probability, and denote the transition rates in the extended and reduced models by $W_{(i,{\bf X}) \to (j,{\bf Y})}$ and $k_{i \to j}$ respectively.
  
  Again, we assume that there is no cycle in the extended model and therefore the detailed balance is satisfied.  We also assume that
\begin{eqnarray}
F({\bf X},i,j) \equiv \sum_{\bf Y} W_{(i,{\bf X}) \to (j,{\bf Y})}   
\end{eqnarray}
is a slowly varying function of $X$. That is, there is a large number $N \gg 1$ such that $F({\bf X},i,j)$ does not deviate significantly from its initial value $F({\bf X}_0,i,j)$ if $|{\bf X}| \ll N$:
\begin{eqnarray}
F({\bf X},i,j) \simeq k_{i \to j} \equiv F({\bf X_0},i,j)  \quad ({\rm for}\ \ |{\bf X}| \ll N). \label{simp2}
\end{eqnarray}
 Then, the transition rates of the driven system are obtained by summing over the states of the driving agent~(Appendix \ref{trcnd}):
\begin{eqnarray}
&&w(i \to j; t) \equiv \Delta t^{-1}(q(i \to j; t) - \delta_{i,j} ) \nonumber\\
&=& \Delta t^{-1} \frac{1}{\pi_i(t)} \sum_{{\bf X},{\bf Y}}\left( {\rm Pr}(i,{\bf X},t; j,{\bf Y},t + \Delta t) - {\rm Pr}(i,{\bf X},t; j,{\bf Y},t) \right)\nonumber\\
&=& \frac{1}{\pi_i(t)}\sum_{{\bf X},{\bf Y}} \Pi_{(i,{\bf X})}(t) W_{(i,{\bf X}) \to (j,{\bf Y})}
\nonumber\\
&\simeq& k_{i \to j} \frac{\sum_{{\bf X}} \Pi_{(i,{\bf X})}(t)}{\pi_i(t)}  \simeq	k_{i \to j}
	 \label{reduced2}
\end{eqnarray}
for $t \ll  t_1 \equiv N/k$, with $k$  being the typical size of $k_{i \to j}$. Again, the summation of ${\bf X}$ in the numerator of the last line excludes the states at the boundary of the Markov network, whose effect is negligible at early times, leading to the final approximation. Although the definition of the transition rate for discrete-time model has been used, the final result does not depend on $\Delta t$, and Eq.(\ref{reduced2}) can be used for both discrete-time and continuous-time models.

From here on, let us refer to the time regime $t \ll t_1$, where the system can be described by a time-homogeneous cyclic Markov model without detailed balance, as the cyclic regime.

\section{Entropy production}
\subsection{Continuous time}
The formalism presented in this work clarifies the notion of entropy production~\cite{QHx1,QHx2,hi12,zmh,hi11,jap,hi12a,hi13a,seif1,seif2,sch,AG,tom1,tom2,zia1,zia2} for the case of a Markov process without detailed balance. The entropy production is connected with time-irreversibility via the fluctuation theorem~\cite{evan,gc,kurc,lebo,crooks,maes,seif1,seif2,gj,episto,no,park,kaw,hi13}  . For a continuous-time Markov process described by the transition rates $k_{i \to j}$, the entropy production of the whole closed system has been defined as
\begin{eqnarray}
\Sigma  \equiv \sum_{i,j} \pi_i(t) k_{i \to j} \log \left(\frac{ \pi_i(t) k_{i \to j}}{\pi_j(t)  k_{j \to i}} \right),\label{sch}
\end{eqnarray}
where the Boltzmann constant has been set to unity\footnote{The base of the $\log$ function will be kept arbitrary, because the results do not depend on it.}. This formula for the entropy production was  originally proposed by Schnakenberg \cite{sch}, and is widely used nowadays~\cite{QHx1,QHx2,hi12,zmh,hi11,jap,hi12a,hi13a,seif1,seif2,AG,tom1,tom2,zia1,zia2}.
It can be shown that $\Sigma  \ge 0$ for any Markov process~(Appendix \ref{entcyc}, \ref{sigcyc}). The Schnakenberg entropy production $\Sigma$ is explicitly expressed as the time-derivative of an entropy in the case of a Markov model that satisfies Kolmogorov's criterion.  From the detailed balance condition in Eq.(\ref{db}), we obtain
\begin{eqnarray}
\Sigma &=& \sum_{i,j} \pi_i(t) k_{i \to j} \log \left(\frac{ \pi_i(t) \pi^{\rm eq}_j}{\pi_j(t)  \pi^{\rm eq}_i} \right)\nonumber\\
&=& -\sum_{i,j} \left( \pi_j(t) k_{j \to i} - \pi_i(t) k_{i \to j} \right) \log \left(\frac{ \pi_i(t) }{\pi^{\rm eq}_i} \right)\nonumber\\
&=& -\sum_{i} \dot \pi_i(t) \log \left(\frac{ \pi_i(t) }{\pi^{\rm eq}_i} \right)\nonumber\\
&=& -\frac{d}{dt}\left[ \sum_{i} \pi_i(t) \log \left(\frac{ \pi_i(t) }{\pi^{\rm eq}_i} \right) \right],\label{sig1}
\end{eqnarray}
where the condition $\sum_i \dot \pi_i(t) = 0$ was used to derive the last line. Therefore, we find that $\Sigma = \dot S_{\rm closed}$, where 
\begin{equation}
S_{\rm closed} = -\sum_i \pi_i(t) \log \left(\frac{\pi_i(t)}{\pi^{\rm eq}_i}\right) \label{tot}
\end{equation}
is the entropy of the whole closed system. 
It takes the form of the negative of the relative entropy, also called the Kullback-Leibler divergence~\cite{KL}. The Kullback-Leibler divergence $D_{\rm KL}(P||Q)$ measures the distance of a probability distribution $P(i)$ from a given distribution $Q(i)$, which is defined as
\begin{equation}
D_{\rm KL}(P||Q)=\sum_i P(i) \log \frac{P(i)}{Q(i)}.
\end{equation}
 Because of the sign flip, $S_{\rm closed}$  measures the similarity of the distribution $\pi(t)$ to $\pi^{\rm eq}_i$. Therefore, we may say that $S_{\rm closed}$ is a nondecreasing function of time because $\pi$ converges to $\pi^{\rm eq}_i$ as time proceeds. However, $\dot S_{\rm closed} \ge 0$ even when $\lim_{t \to \infty} \pi_i(t) \ne \pi^{\rm eq}_i$~(Appendix \ref{entcyc}). The physical interpretation of $S_{\rm closed}$ is very clear. At the equilibrium, a closed system has an equal probability to be in each microstate that is consistent with the given constraints~\cite{lee,tolman} (Appendix \ref{shanlee}). Consequently, the equilibrium probability $\pi^{\rm eq}_i$ is proportional to the number of such microstates corresponding to the state $i$, denoted as $\Omega_i$~\cite{lee} (Appendix \ref{shanlee}):
 \begin{equation}
 \pi^{\rm eq}_i \propto \Omega_i = B^{S_i}, \label{eqd}
 \end{equation}
 where $S_i \equiv \log \Omega_i$ is the Boltzmann entropy corresponding to the state $i$, and $B$ is the base of the log function. From Eq.(\ref{eqd}), we have
\begin{equation}
 S_{\rm closed} =  -\sum_i \pi_i(t) \log \pi_i(t) + \sum_i \pi_i(t) S_i(t)    + {\rm const}. \label{sysmed}
\end{equation}
The first term in Eq.(\ref{sysmed}), called the Shannon entropy~\cite{rmp,shannon}, results from the uncertainty of the slow variable $i$. The second term, the average Boltzmann entropy, is due to the uncertainty of the remaining fast degree of freedom that is in instantaneous local equilibrium~(Appendix \ref{shanlee}).

For a cyclic Markov model without detailed balance, the entropy in Eq.(\ref{tot}) is still a nondecreasing function of time~(Appendix \ref{entcyc}). We now denote it as
\begin{equation}
S_{\rm cyc} \equiv -\sum_i \pi_i(t) \log \left(\frac{\pi_i(t)}{\pi^{\rm st}_i}\right), \label{scy}
\end{equation} 
with $\pi^{\rm st}_i$ being the stationary distribution without detailed balance. However, $S_{\rm cyc}$ does not lend itself to a clear physical interpretation as Eq.(\ref{sysmed}). Furthermore, although the fact that $\Sigma \ge 0$ remains true regardless of the detailed balance, $\Sigma$ is no longer equal to $\dot S_{\rm cyc}$. In fact, it has been shown that $\Sigma > \dot S_{\rm cyc}$\cite{hg,gq} in the absence of detailed balance (Appendix \ref{sigcyc}). No analytic expression for the entropy component, whose time derivative is $\Sigma$, has been constructed so far.

In this section, it will be shown that by embedding the cyclic Markov model into a larger Markov model with detailed balance that explicitly includes the degrees of freedom for the drivers, the Schnakenberg entropy production $\Sigma$ can be explicitly expressed as a time derivative of an entropy component under an appropriate condition, justifying its identity as an entropy production. The total entropy of the extended model is simply obtained from Eq.(\ref{tot}) by making the replacements $\pi_i(t) \to \Pi_{(i,{\bf X})}(t)$ and $\pi^{\rm eq}_i \to \Pi^{\rm eq}_{(i,{\bf X})}$: 
\begin{equation}
S_{\rm tot}\equiv -\sum_{i,{\bf X}}  \Pi_{(i,{\bf X})}(t)\log \frac{\Pi_{(i,{\bf X})}(t)}{\Pi^{\rm eq}_{(i,{\bf X})}}
. \label{clex}
\end{equation}
By performing the decomposition 
\begin{equation}
\Pi_{(i,{\bf X})}(t) = \pi_i (t) \Pi_{({\bf X}/i)}(t) ,
\end{equation}
where  $\pi_i(t) \equiv \sum_{\bf X} \Pi_{(i,{\bf X})}(t)$ is the marginal probability and $\Pi_{({\bf X}/i)}(t) \equiv \Pi_{(i,{\bf X})}(t)/\pi_i(t)$ is the conditional probability, $S_{\rm tot}$  is now decomposed as
\begin{equation}
S_{\rm tot} =  -\sum_i \pi_i(t) \log \pi_i(t) + \sum_{i,{\bf X}} \Pi_{(i,{\bf X})}(t) \log \Pi^{\rm eq}_{(i,{\bf X})} -\sum_{i,{\bf X}} \Pi_{(i,{\bf X})}(t)  \log \Pi_{({\bf X}/i)}(t). \label{dcmp1}
\end{equation}
The first term, the Shannon entropy of the driven system
\begin{equation}
S_{\rm shan} = -\sum_i \pi_i(t) \log \pi_i(t), \label{shan1}
\end{equation}
results from the uncertainty of the state $i$ of the driven system.
The third term, the hidden entropy
\begin{equation}
S_{\rm hid} =  -\sum_{i,{\bf X}} \Pi_{(i,{\bf X})}(t)  \log \Pi_{({\bf X}/i)}(t), \label{hid1}
\end{equation}
originates from the uncertainty of the driver degrees of freedom ${\bf X}$ for a given value of $i$.
Finally, the second term, the average Boltzmann entropy
\begin{equation}
S_{\rm bol} =    \sum_{i,{\bf X}} \Pi_{(i,{\bf X})}(t) \log \Pi^{\rm eq}_{(i,{\bf X})}= \sum_{i,{\bf X}} \Pi_{(i,{\bf X})}(t) S_{(i,{\bf X})},\label{abol}
\end{equation}
comes from the uncertainty of the remaining fast degrees of freedom for given $(i,{\bf X})$. Because fast degrees of freedom are locally equilibrated, the corresponding indices do not appear explicitly in  Eq.(\ref{abol})~\cite{lee}(Appendix \ref{shanlee}). 

It is straightforward to show that (Appendix \ref{totsig})
\begin{equation}
\dot S_{\rm tot} - \dot S_{\rm hid} =\dot S_{\rm shan} + \dot S_{\rm bol} = \Sigma_{\rm exact} \label{entpr}
\end{equation}
where 
\begin{eqnarray}
\Sigma_{\rm exact} &\equiv&  \sum_{i,j,{\bf X},{\bf Y}}   \Pi_{(i,{\bf X})}(t) W_{(i,{\bf X}) \to (j,{\bf Y})}  \log \frac{\pi_i (t) W_{(i,{\bf X}) \to (j,{\bf Y})} }{\pi_j (t) W_{(j,{\bf Y}) \to (i,{\bf X})} }.
\label{exent} 
\end{eqnarray}

We now assume that in the cyclic regime where Eq.(\ref{simp2}) holds, the ratio $ W_{(i,{\bf X}) \to (j,{\bf Y})} / W_{(j,{\bf Y}) \to (i,{\bf X})}$ is also determined solely by the indices $i$ and $j$. That is, we assume that
\begin{equation}
\frac{W_{(i,{\bf X}) \to (j,{\bf Y})}}{W_{(j,{\bf Y}) \to (i,{\bf X})}} \simeq r_{i j},\label{condom}
\end{equation}
which can be rewritten as
\begin{equation}
W_{(i,{\bf X}) \to (j,{\bf Y})}= r_{i j} W_{(j,{\bf Y}) \to (i,{\bf X})}. \label{111}
\end{equation}
By summing Eq.(\ref{111}) over ${\bf X}$ and ${\bf Y}$, we get $r_{i j}=k_{i \to j}/k_{j \to i}$. Therefore, the condition Eq.(\ref{condom}) can be rewritten as
\begin{equation}
\frac{W_{(i,{\bf X}) \to (j,{\bf Y})}}{W_{(j,{\bf Y}) \to (i,{\bf X})}} \simeq \frac{k_{i \to j}}{k_{j \to i}}.\label{condom2}
\end{equation}

Under the conditions Eq.(\ref{simp2}) and Eq.(\ref{condom2}), 
 $\Sigma_{\rm exact}$ is approximated as
\begin{eqnarray}
\Sigma_{\rm exact} &\simeq& \sum_{i,j,{\bf X},{\bf Y}}   \Pi_{(i,{\bf X})}(t) W_{(i,{\bf X}) \to (j,{\bf Y})}  \log \frac{\pi_i (t) k_{i\to j} }{\pi_j (t) k_{j \to i} } \nonumber\\
&=& \sum_{i,j,{\bf X}}   \Pi_{(i,{\bf X})}(t) F({\bf X},i,j) \log \frac{\pi_i (t) k_{i\to j} }{\pi_j (t) k_{j \to i} } \nonumber\\
&\simeq& \sum_{i,j}  \Pi_{(i,{\bf X})}(t) k_{i \to j} \log \frac{\pi_{i}(t) k_{i \to j}}{\pi_{j}(t) k_{j \to i}} = \Sigma.
\end{eqnarray}
and we get
\begin{equation}
\dot S_{\rm tot} \simeq  \Sigma +  \dot S_{\rm hid}.
\end{equation}
The hidden entropy $S_{\rm hid}$ is a newly identified entropy component that cannot be expressed in terms of the parameters of the reduced model. Hidden entropy production has been discussed previously\cite{kaw,epi,chun}, but the analytic expression for $S_{\rm hid}$ itself has not been derived up to the present. I also derived the condition Eq.(\ref{condom2}), required in addition to Eq.(\ref{simp2}), for the Schnakenberg entropy production $\Sigma$ to be equal to $\dot S_{\rm tot}-\dot S_{\rm hid}$. If these conditions are not satisfied, then $\Sigma$ should be replaced by the exact form $\Sigma_{\rm exact}$ in Eq.(\ref{exent}).

\subsection{Discrete time}
Even for a discrete Markov jump process, $S_{\rm closed}$ and $S_{\rm cyc}$, defined by Eq. (\ref{tot}) and Eq. (\ref{scy}) respectively, are nondecreasing functions of time~(Appendix \ref{entcyc})\cite{morimoto}. 
The discrete-time counterpart of $\dot S_{\rm closed}$ is~(Appendix \ref{rem}):
\begin{eqnarray}
\frac{\Delta S_{\rm closed}}{\Delta t} &\equiv& \left[ S_{\rm closed}(t+\Delta t) - S_{\rm closed}(t)\right]\frac{1}{\Delta t} \nonumber\\
&=& {\Delta t}^{-1} \sum_{i,j} \pi_i(t) p_{i \to j}   \log \left(\frac{\pi_i(t) p_{i \to j}}{\pi_j(t +\Delta t) p_{j \to i}}\right) \nonumber\\
&=& \sum_{i,j} \pi_i(t) k_{i \to j}   \log \left(\frac{\pi_i(t) k_{i \to j}}{\pi_j(t +\Delta t) k_{j \to i}}\right) + {\Delta t}^{-1} \sum_{i} \pi_i(t)  \log \left(\frac{\pi_i(t) }{\pi_i(t +\Delta t) }\right)
.\label{entpd}
\end{eqnarray}
 Therefore, it is reasonable to generalize Eq.(\ref{entpd}) to a Markov model without detailed balance and define the Schnakenberg entropy production $\Sigma$ as
\begin{eqnarray}
\Sigma &\equiv& {\Delta t}^{-1} \sum_{i,j} \pi_i(t) p_{i \to j}   \log \left(\frac{\pi_i(t) p_{i \to j}}{\pi_j(t +\Delta t) p_{j \to i}}\right)\nonumber\\
&=& \sum_{i,j} \pi_i(t) k_{i \to j}   \log \left(\frac{\pi_i(t) k_{i \to j}}{\pi_j(t +\Delta t) k_{j \to i}}\right) + {\Delta t}^{-1} \sum_{i} \pi_i(t)   \log \left(\frac{\pi_i(t) }{\pi_i(t +\Delta t) }\right)
\label{schd}
\end{eqnarray}
for the case of a discrete-time model.
As in the case of continuous-time models, $\Sigma > {\Delta S_{\rm cyc}}/{\Delta t} \ge 0$ for discrete-time models in the absence of detailed balance~(Appendix \ref{sigcyc}). 

We now define $\Sigma_{\rm exact}$ as 
\begin{eqnarray}
\Sigma_{\rm exact} &\equiv& \sum_{i,j,{\bf X}}  \Pi_{(i,{\bf X})}(t) W_{(i,{\bf X}) \to (j,{\bf Y})} \log \frac{\pi_{i}(t) W_{(i,{\bf X}) \to (j,{\bf Y})}}{\pi_{j}(t+ \Delta t) W_{(j,{\bf Y}) \to (i,{\bf X})}}\nonumber\\
&&+ \Delta t^{-1} \sum_{i } \pi_{i}(t)  \log \frac{\pi_{i}(t) }{\pi_{i}(t+ \Delta t)}, \label{exd}
\end{eqnarray}
which is the discrete-time counterpart of Eq.(\ref{exent}). It is then straightforward to show that~(Appendix \ref{totsig}) 
\begin{equation}
\Sigma_{\rm exact} = \frac{\Delta S_{\rm tot}}{\Delta t}- \frac{\Delta S_{\rm hid}}{\Delta t} = \frac{\Delta S_{\rm shan}}{\Delta t}+ \frac{\Delta S_{\rm bol}}{\Delta t},\label{fin1}
\end{equation}
with the same definitions of $S_{\rm tot}$, $S_{\rm hid}$, $S_{\rm shan}$, and $S_{\rm bol}$ (Eqs.(\ref{clex}),(\ref{hid1}),(\ref{shan1}), and (\ref{abol})) as in the case of continuous time. 

In the regime where the conditions Eq.(\ref{simp2}) and Eq.(\ref{condom2}) are satisfied, we have
\begin{eqnarray}
\Sigma_{\rm exact} &\simeq& \sum_{i,j}  \pi_{i}(t) k_{i \to j} \log \frac{\pi_{i}(t) k_{i \to j}}{\pi_{j}(t+ \Delta t) k_{j \to i}} \nonumber\\
&+&  \Delta t^{-1} \sum_{i } \pi_{i}(t)  \log \frac{\pi_{i}(t) }{\pi_{i}(t+ \Delta t)}\nonumber\\
&=& \Sigma,
\end{eqnarray} 
and therefore
\begin{equation}
\frac{\Delta S_{\rm tot}}{\Delta t}  \simeq \Sigma + \frac{\Delta S_{\rm hid}}{\Delta t} .
\end{equation}
\subsection{Examples with $\Sigma_{\rm exact} \simeq \Sigma$}
Eq.(\ref{simp2}) and Eq.(\ref{condom2}) are the crucial assumptions for   $\Sigma_{\rm exact}=\dot S_{\rm tot} - \dot S_{\rm hid}$ to be approximated by the Schnakenberg entropy production $\Sigma$. We discuss a couple of examples where these two conditions are satisfied. 

First, let us assume that the change of ${\bf X}$, $\Delta {\bf X}$, is uniquely determined by the states of the driven system before and after the transition. We write this as $\Delta {\bf X}(i \to j)$, where $i$ and $j$ denote the states before and after the transition, respectively. Then, the transition rate takes the form
\begin{equation}
W_{(i,{\bf X}) \to (j,{\bf Y})} = g(i \to j; {\bf X}) \delta({\bf Y}-{\bf X},\Delta {\bf X}(i \to j)). \label{simp3}
\end{equation}
We also assume that  
\begin{equation}
\Delta {\bf X}(i \to j) = -\Delta {\bf X}(j \to i). \label{another}
\end{equation}
 The transition rates in the example considered earlier, namely the ATP-driven biochemical cycle with a fixed total number of ATP and ADP, takes the form of Eq. (\ref{simp3}) with Eq. (\ref{another}). If $g(i \to j; {\bf X})$ is a slowly-varying function, then
\begin{equation}
W_{(i,{\bf X}) \to (j,{\bf Y})} \simeq k_{i \to j} \delta({\bf Y}-{\bf X},\Delta {\bf X}(i \to j))  \label{simp4}
\end{equation}
at early times.
It is easy to see that Eq. (\ref{simp4}) and Eq. (\ref{another}) imply that both Eq. (\ref{simp2}) and Eq. (\ref{condom2}) hold.

The change of ${\bf X}$ for a given transition $i \to j$ does not have to be unique in order for the conditions in Eq. (\ref{simp2}) and Eq. (\ref{condom2}) to hold. For example, let us say that the number of driver molecules consumed for the transitions $C \to A$ and $A \to C$ in the three-state model considered earlier are not unique. Then, the transition rates take the form $W_{(C,X) \to (A,X-1)} = k_1 X,\  W_{(A,X-1) \to (C,X)} = \tilde k_1 (N-X+1),\ W_{(C,X) \to (A,X-2)} = k_2 X^2,\ W_{(C,X-2) \to (A,X)} = \tilde k_2 (N-X+2)^2, \cdots$, where $X$ is the number of ATP molecules,  and $N$ is the total combined number of ATP and ADP molecules. Eq. (\ref{simp2}) holds at early times because $X$ does not deviate significantly from its initial value. Eq. (\ref{condom2}) is also satisfied if $k_1/\tilde k_1 = k_2/\tilde k_2 \cdots$, because $X \gg 1$ at early times.

\subsection{Housekeeping entropy}
In a cyclic Markov model without detailed balance, the entropy defined in terms of the nonequilibrium steady state,
\begin{equation}
S_{\rm cyc} \equiv -\sum_i \pi_i(t) \log \frac{\pi_i(t)}{\pi^{\rm st}_i}, \label{cy3}
\end{equation}
is often considered, which increases with time as explained earlier. This motivates us to perform an alternative decomposition 
\begin{eqnarray}
S_{\rm tot} &=&  -\sum_i \pi_i(t) \log \frac{\pi_i(t)}{\pi^{\rm st}_i} + \sum_{i,{\bf X}} \Pi_{(i,{\bf X})}(t) \log \frac{\Pi^{\rm eq}_{(i,{\bf X})}}{\pi^{\rm st}_i} -\sum_{i,{\bf X}} \Pi_{(i,{\bf X})}(t)  \log \Pi_{({\bf X}/i)}(t)\nonumber\\
&=& S_{\rm cyc} + S_{\rm hk} + S_{\rm hid},
 \label{dcmp2}
\end{eqnarray}
where we now define the housekeeping entropy $S_{\rm hk}$ as 
\begin{equation}
S_{\rm hk} \equiv \sum_{i,{\bf X}} \Pi_{(i,{\bf X})}(t) \log \frac{\Pi^{\rm eq}_{(i,{\bf X})}}{\pi^{\rm st}_i}.   \label{hk2}
\end{equation} 
It is easy to see the equivalence of Eq. (\ref{dcmp2}) to Eq. (\ref{dcmp1}), because $\pi^{\rm st}_i$ in the first and the second terms of Eq. (\ref{dcmp2}) cancel with each other, leading to Eq. (\ref{dcmp1}). We then see that
\begin{equation}
\Sigma_{\rm exact} - \dot S_{\rm cyc} = \dot S_{\rm hk}
\end{equation}
for a continuous-time model.
In the regime where Eq. (\ref{simp2}) and Eq. (\ref{condom2}) are valid, we get
\begin{equation}
\Sigma - \dot S_{\rm cyc} \simeq \dot S_{\rm hk}.
\end{equation}
It has been argued that even after the steady-state has been reached in a cyclic Markov model, where $\dot S_{\rm cyc} \simeq 0$, heat should be constantly generated in order to maintain the steady state,  called the housekeeping heat~\cite{hg,gq,op,hs,sp}. Clearly, the generation of such heat is proportional to the production of an entropy component. Eq. (\ref{hk2}) is the analytic formula for this entropy component, and was accordingly termed the house-keeping entropy. Details on the relations between $S_{\rm hk}$ and the housekeeping heat are given in Appendices \ref{open} and \ref{lang}.

Both $S_{\rm cyc}$ and $S_{\rm hk}$ are expressed in terms of the relative entropy, and we see that $S_{\rm cyc}$ in Eq. (\ref{cy3}) measures the similarity of the distribution $\pi(t)$ to the quasi-steady state $\pi^{\rm st}_i$. Also, we see that $S_{\rm hk}$ in Eq. (\ref{hk2}) measures the tendency of $\Pi_{(i,\rm X)}$ to move away from the quasi-steady state and approach the true equilibrium $\Pi^{\rm eq}_{(i,{\bf X})}$.  
  
 Note that from the viewpoint of the extended model, $\pi^{\rm st}_i$ is just a quasi-steady state. Therefore, in contrast to equilibrium distribution $\Pi^{\rm eq}_{(i,{\bf X})} \propto \Omega_{(i,{\bf X})}$, which is expressed in terms of the number of microstates $\Omega_{(i,{\bf X})}$ for a given $(i,{\bf X})$, the quasi-steady state $\pi^{\rm st}_i$ is just an artefact of the dynamics, and does not seem to have a microscopic interpretation as in the case of $\Pi^{\rm eq}_{(i,{\bf X})}$. Because the alternative decomposition in Eq. (\ref{dcmp2}) is defined in terms of $\pi^{\rm st}_i$, $S_{\rm cyc}$ and $S_{\rm hk}$ do not lend themselves to clear interpretations as uncertainties of some degrees of freedom, in contrast to $S_{\rm shan}$ and $S_{\rm bol}$. Although we considered a continuous-time model in this section, all of the results given here are valid for a discrete-time model if we make the replacements $\dot S_{\rm cyc} \to \Delta S_{\rm cyc}/\Delta t$ and $\dot S_{\rm hk} \to \Delta S_{\rm hk}/\Delta t$. The explicit connections of $S_{\rm hk}$ and $S_{\rm hid}$ to the quantities considered in previous literature are provided in Appendices \ref{open}, \ref{lang}, and \ref{adna}.

\subsection{Behavior of various entropy components}
Let us summarize the general behavior of various entropy components. We will use the notation for the continuous-time Markov process. The result for the discrete-time Markov process is obtained by simply replacing $\dot S$ with $\Delta S/\Delta t$. We assume that the condition Eq.(\ref{condom2}) holds in the cyclic regime so that $\Sigma_{\rm exact} \simeq \Sigma$. From the results in the literature for cyclic models\cite{hg,gq,morimoto,zmh}, we already know that $\dot S_{\rm cyc} \ge 0$ and $\Sigma-\dot S_{\rm cyc} \ge 0$ in the cyclic regime, and therefore $\dot S_{\rm hk} \ge 0$~(Appendix \ref{entcyc}, \ref{sigcyc}). Once the system reaches the quasi-steady state, $\dot S_{\rm cyc} \simeq 0$.   By embedding the cyclic Markov model into a larger model, it can be shown that $\dot S_{\rm hid} \ge 0$ in the cyclic regime~(Appendix \ref{dhid}). During the transition from the quasi-steady state to true equilibrium, we have $\dot S_{\rm cyc} \le 0$, because the distribution diverges from the quasi-steady state. However, we have $\dot S_{\rm hk} + \dot S_{\rm hid} \ge 0$ because $\dot S_{\rm tot} \ge 0$. All entropy components will reach constant values after the system reaches equilibrium. Various entropy components for the previous cyclic three-state discrete-time Markov jump process are shown in Figure 3, where this general behavior is confirmed. 

\section{Conclusion}
  It has been shown that for any time-homogeneous Markov process that violates Kolmogorov's criterion, one can always find a larger Markov process that satisfies the criterion, where the degrees of freedom ${\bf X}$ for the driving agent are explicitly included. The original Markov model is then recovered as an approximation at early times after eliminating ${\bf X}$. The nonequilibrium steady state of the original model is then found to be a quasi-steady state.  
  
   An important contribution of the current work is that by extending the cyclic model to a model with detailed balance, we indeed find analytic expressions for $S_{\rm hk}$ and $S_{\rm hid}$ that satisfy $\dot S_{\rm hk}  = \Sigma - \dot S_{\rm cyc}$ and $\dot S_{\rm hid} = \dot S_{\rm tot} - \Sigma$ in the cyclic regime. Here, $S_{\rm hk}$ itself cannot be expressed with parameters in the cyclic Markov model,  but its derivative $\Sigma - \dot S_{\rm cyc}$ can. Furthermore, neither $S_{\rm hid}$ nor its derivative can be expressed with parameters of the cyclic Markov model. That is, they are completely hidden in the cyclic Markov model description.
  
  The current formalism is very general and can be applied to any closed system. Such a closed system includes, but is not limited to, an open system and an infinitely large heat bath in contact with each other~(Appendix \ref{open}). Although we assumed that the state index $i$ is discrete in this work, the extension of the current formalism to a Markov process with a continuous index such as Langevin dynamics~\cite{seif1,seif2,hs,sp,hy2}, is straightforward~(Appendix \ref{lang}). Note that the construction of the extended system is by no means unique. The situation is analogous to that of the canonical ensemble of an equilibrium system, where the properties of the system depend only on the temperature of the heat bath and microscopic details of the bath are arbitrary.
   
\section{Acknowledgement} 
This work was supported by the National Research Foundation of Korea, funded by the Ministry of Education, Science, and Technology (NRF-2017R1D1A1B03031344). The author thanks Changbong Hyeon for useful suggestions.



\appendix
\section{Definition of transition rate for a general dynamic system}\label{trcnd}
In {\it any} stochastic process with discrete time, the transition probability $q(i \to j; t)$  is defined as the conditional probability that the state of the system is $j$ at time $t+\Delta t$, given that the state is $i$ at time $t$:
\begin{equation}
q(i \to j;t) \equiv \frac{{\rm Pr}(i,t;j,t+\Delta t)}{\pi(i,t)} \label{cond}
\end{equation}
where $\pi_i(t)$ is the probability that the state of the system is $i$ at time $t$, and ${\rm Pr}(i,t; j,t')$ is the joint probability that the state of the system is $i$ at time $t$ and $j$ at time $t'$. By multiplying both sides of Eq.(\ref{cond}) by $\pi_i(t)$ and summing over $i$, we obtain
\begin{equation}
\sum_i \pi_i(t) q (i \to j;t) = \sum_i {\rm Pr}(i,t;j,t+\Delta t) = \pi_j(t+\Delta t). \label{dmagain}
\end{equation}  
Note that for a general non-Markov process, the transition probability $q(i \to j; t)$ is not a time-independent constant, and its value may depend on the previous history of the system. Only for the special case when $q(i \to j;t)$ is a time-independent constant $p_{i \to j}$ do we recover the dynamical equation for a time-homogeneous Markov jump process, Eq.(\ref{markov2}).

Analogous to the procedure in the Markov model, the transition rate $w(i \to j; t)$ is also defined as
\begin{equation}
w(i \to j; t) \equiv (\Delta t)^{-1}(q(i \to j;t) - \delta_{i,j}) \label{rateagain}
\end{equation}
and Eq.(\ref{dmagain}) can be rewritten as
\begin{equation}
\frac{\Delta \pi_i (t)}{\Delta t} = \sum_j \pi_j(t) w(j \to i; t). \label{dmagain2}
\end{equation}
Note that by substituting Eq.(\ref{cond}) into Eq.(\ref{rateagain}), the transition rate $w(i \to j, t)$ can be written as
\begin{eqnarray}
w(i \to j; t) &=& (\Delta t)^{-1} \left(\frac{{\rm Pr}(i,t;j,t+\Delta t)}{\pi(i,t)}-\delta_{i,j}\right)= (\Delta t)^{-1} \frac{{\rm Pr}(i,t;j,t+\Delta t)-\delta_{i,j} \pi(i,t)}{\pi(i,t)}\nonumber\\
&=&  (\Delta t)^{-1} \frac{{\rm Pr}(i,t;j,t+\Delta t)-{\rm Pr}(i,t;j,t)}{\pi_i(t)}. \label{alk}
\end{eqnarray}
where we used the fact that ${\rm Pr}(i,t;j,t)=\delta_{i,j} \pi_i(t)$. 

A stochastic process with continuous time is obtained from a discrete-time model by taking the limit of $\Delta t \to 0$. Then, Eq.(\ref{dmagain2}) reduces to
\begin{equation}
\dot \pi_i (t) = \sum_j \pi_j(t) w(j \to i; t),
\end{equation}
where Eq.(\ref{alk}) now becomes
\begin{equation}
w(i \to j; t) = \frac{1}{\pi_i(t)}\frac{d}{dt'}{\rm Pr}(i,t;j,t')|_{t'=t}.
\end{equation}

Again, the transition rate $w(j \to i; t)$ may in general depend on the previous history of the system. In the special case that $w(j \to i; t)$ is a time-independent constant $k_{i \to j}$, we recover the time-homogeneous master equation Eq.(\ref{markov}).

\section{Analytic form of the equilibrium solution of the extended three-state model}\label{eqsol}
Here, the analytic form of the equilibrium distribution of the extended three-state model (Fig.1(b,c)) is presented.
By defining $X$ as a serial coordinate labelling the states of the extended model (Fig 1(b)), we obtain a simple relation between the label $i$ of the driven system and $X$:
\begin{equation}
i=X \mod 3,
\end{equation}
where $i=1,0,2$ correspond to the states A,B,C, if we assume that the driven system is at state B for $X=0$. Because $i$ is uniquely determined by $X$, we will write the equilibrium distribution $\Pi^{\rm eq}_{i,X}$ simply as $\Pi^{\rm eq}_{X}$. The equilibrium distribution of model 1 can be obtained analytically using the following detailed balance conditions:
\begin{eqnarray}
\Pi^{\rm eq}_{3m+1}a &=& \Pi^{\rm eq}_{3m}\alpha \nonumber\\
\Pi^{\rm eq}_{3m+2}c &=& \Pi^{\rm eq}_{3m+1}\gamma \nonumber\\
\Pi^{\rm eq}_{3m+3}b &=& \Pi^{\rm eq}_{3m+2}\beta, \label{eqde}
\end{eqnarray}
where $m$ is an integer.
Because the network topology is linear, one can use Eq.(\ref{eqde}) as a recursion relation in order to express $\Pi^{\rm eq}_{X}$ in terms of the state 
$\Pi^{\rm eq}_{0}$:
\begin{eqnarray}
\Pi^{\rm eq}_{3m} &=& \left(\frac{\alpha\beta\gamma}{abc}\right)^m\Pi^{\rm eq}_{0} \nonumber\\
\Pi^{\rm eq}_{3m+1} &=& \left(\frac{\alpha\beta\gamma}{abc}\right)^m\frac{\alpha}{a}\Pi^{\rm eq}_{0} \nonumber\\
\Pi^{\rm eq}_{3m+2} &=& \left(\frac{\alpha\beta\gamma}{abc}\right)^m\frac{\gamma \alpha}{ca}\Pi^{\rm eq}_{0} . \label{eqdi}
\end{eqnarray}
The constant $\Pi^{\rm eq}_{0}$ is obtained from the normalization condition $\sum_{X=0}^N \Pi^{\rm eq}_X = 1$ as
\begin{equation}
\Pi^{\rm eq}_{0} = \left[ \frac{(1- \left(\frac{\alpha\beta\gamma}{abc}\right)^{N/3})(1+ \frac{\alpha}{a}+\frac{\gamma \alpha}{ca})}{1- \left(\frac{\alpha\beta\gamma}{abc}\right)} + \left(\frac{\alpha\beta\gamma}{abc}\right)^{N/3} \right]^{-1},\label{pino}
\end{equation}
where the formula for the summation of the geometric series was used.

For model 2, the detailed balance condition is the same as in Eq.(\ref{eqde}) except that the second line of Eq.(\ref{eqde})is modified to
\begin{equation}
\Pi^{\rm eq}_{3m+2} c(m+1) = \Pi^{\rm eq}_{3m+1} \gamma (N/3-m).
\end{equation}
The solution of the recurrence relation is now
\begin{eqnarray}
\Pi^{\rm eq}_{3m} &=& \left(\frac{\alpha\beta\gamma}{abc}\right)^m\frac{(N/3)!}{m! (N/3-m)!}\Pi^{\rm eq}_{0} \nonumber\\
\Pi^{\rm eq}_{3m+1} &=& \left(\frac{\alpha\beta\gamma}{abc}\right)^m\frac{\alpha}{a}\frac{(N/3)!}{m! (N/3-m)!}\Pi^{\rm eq}_{0} \nonumber\\
\Pi^{\rm eq}_{3m+2} &=& \left(\frac{\alpha\beta\gamma}{abc}\right)^{m+1} \frac{b}{\beta}\frac{(N/3)!}{(m+1)! (N/3-m-1)!}\Pi^{\rm eq}_{0} . \label{eqdi3}
\end{eqnarray}
Again, the constant $\Pi^{\rm eq}_{0}$ is obtained from the normalization condition $\sum_{X=0}^N \Pi^{\rm eq}_X = 1$ as
\begin{equation}
\Pi^{\rm eq}_{0}=\left[\left(1 +\frac{\alpha\beta\gamma}{abc}\right)^{N/3}\left(1+ \frac{\alpha}{a}+\frac{b}{\beta}\right)  -  \left(\frac{\alpha\beta\gamma}{abc}\right)^{N/3}\frac{\alpha}{a}   - \frac{b}{\beta}\right]^{-1}, \label{pino2}
\end{equation}
where the formula for the binomial expansion was used.

The equilibrium distribution is shown in log scale as a function of $X$ in Figure \ref{equi}, where the parameters were set to the values used in the numerical computation. As expected, $\Pi^{\rm eq}_X$ for model 1 is a decreasing function of $X$, because the rate for $X \to X+1$ is always less than that for $X+1 \to X$. In contrast, this is not the case for model 2, where $\Pi^{\rm eq}_X$ has a peak around $X=1000$. This is because for small $X$, the small number of ATP discourages the forward reaction $X+1 \to X$, and the large number of ADP encourages the reverse reaction $X \to X+1$, pushing the system away from $X=0$. 

The equilibrium probability distribution $\pi^{\rm eq}_i$ for model 1 can be obtained by summing the expressions in (\ref{eqdi}) over $m$. The summation range is $0 \le m \le N$ for $\Pi^{\rm eq}_{3m}$ and $0 \le m \le N-1$ for the others. However, for the parameter values used in the numerical computation, we have $({\alpha\beta\gamma}/{abc})^{N/3}=(1/2)^{1000}$, which is negligible for all practical purposes. Therefore, $\sum_{m=0}^{N/3}({\alpha\beta\gamma}/{abc})^m \simeq \sum_{m=0}^{N/3-1}({\alpha\beta\gamma}/{abc})^m$, and we get $\pi^{\rm eq}_0 : \pi^{\rm eq}_1 : \pi^{\rm eq}_2  = 1:\alpha/a: \gamma\alpha/ca = 1:1:0.5$, leading to $(\pi^{\rm eq}_A, \pi^{\rm eq}_B, \pi^{\rm eq}_C) = (2/5,2/5,1/5)$. Similarly, $\pi^{\rm eq}_i$ for model 2 is obtained by summing the expressions in (\ref{eqdi3}) over $m$, and we get $\pi^{\rm eq}_0 : \pi^{\rm eq}_1 : \pi^{\rm eq}_2  = 1:\alpha/a: b/\beta = 1:1:1$, leading to $(\pi^{\rm eq}_A, \pi^{\rm eq}_B, \pi^{\rm eq}_C) = (1/3,1/3,1/3)$. One can check that $\pi_i (t)$ indeed converges to $\pi^{\rm eq}_i$ at late times of the simulation, both for model 1 and model 2 (Figure \ref{steady} (b),(c)).

\section{Proof of $S_{\rm cyc}/\Delta t \ge 0$ or $\dot S_{\rm cyc} \ge 0$ }\label{entcyc}
We prove that the relative entropy
\begin{equation}
S_{\rm cyc} \equiv -\sum_i \pi_i(t) \log \left(\frac{\pi_i(t)}{\pi^{\rm st}_i}\right)
\end{equation}
is a non-decreasing function of time, where $\pi^{\rm st}_i$ the stationary distribution. Regarding the special case of a Markov model with detailed balance, $\pi^{\rm st}_i$ is the equilibrium distribution $\pi^{\rm eq}_i$, and $S_{\rm cyc}$ is denoted as $S_{\rm closed}$. The proof for the discrete-time case has been provided by Morimoto~\cite{morimoto}, which is reproduced below. First, note the definition of a convex function $\phi(x)$, which states that 
\begin{equation}
a \phi (x_1) + (1-a) \phi (x_2) \ge \phi (a x_1 + (1-a) x_2)  
\end{equation}
for any combination of $x_1$, $x_2$, and $a$ with $0 \le a \le 1$. 
As an immediate consequence, for any nonnegative numbers $a_j$ with
\begin{equation}
\sum_j a_j=1,\label{nor}
\end{equation}
 we have
\begin{equation}
\sum_j a_j \phi (x_j) \ge \phi (\sum_j a_j x_j). \label{conv}
\end{equation}
The nonnegative quantities defined by $a_j \equiv \frac{\pi^{\rm st}_j p_{j \to i}}{\pi^{\rm st}_i}$ satisfy the normalization condition in Eq.(\ref{nor}) due to the balance condition in Eq.(\ref{balance}). Therefore, by setting $x_j= \pi_j (t)/\pi^{\rm st}_j$, we obtain 
\begin{equation}
 \sum_j \frac{\pi^{\rm st}_j p_{j \to i}}{\pi^{\rm st}_i} \phi \left(\frac{\pi_j(t)}{\pi^{\rm st}_j}\right) \ge \phi \left(\sum_j \frac{\pi_j(t)  p_{j \to i}}{\pi^{\rm st}_i}\right) = \phi \left( \frac{ \pi_i(t + \Delta t) }{\pi^{\rm st}_i}\right) . \label{conv1}
\end{equation}
Multiplying both sides of Eq.(\ref{conv1}) by $\pi^{\rm st}_i$ and summing over $i$, we obtain
\begin{equation}
 \sum_{i,j} \pi^{\rm st}_j p_{j \to i} \phi \left(\frac{\pi_j(t)}{\pi^{\rm st}_j}\right) = \sum_{j} \pi^{\rm st}_j  \phi \left(\frac{\pi_j(t)}{\pi^{\rm st}_j}\right) \ge  \sum_i \pi^{\rm st}_i \phi \left( \frac{ \pi_i(t + \Delta t) }{\pi^{\rm st}_i}\right). \label{conv2}
\end{equation} 
Introducing the convex function $\phi(x) = x \log x$ into Eq.(\ref{conv2}), we now have
\begin{equation}
 \sum_{j} \pi_j (t) \log \left(\frac{\pi_j(t)}{\pi^{\rm st}_j}\right) \ge  \sum_i \pi_i  (t + \Delta t) \log \left( \frac{ \pi_i(t + \Delta t) }{\pi^{\rm st}_i}\right),\label{eq1}
\end{equation} 
from which we obtain 
\begin{equation}
\frac{\Delta S_{\rm cyc}}{\Delta t} \ge 0.\label{eq2}
\end{equation}

The proof of $\dot S_{\rm cyc} \ge 0$ for a continuous-time model follows straightforwardly from Eq.(\ref{eq2}) by taking the limit $\Delta t \to 0$. Nevertheless, a direct derivation is also presented for the sake of completeness. Note that
\begin{eqnarray}
\dot S_{\rm cyc} &=& -\sum_i \dot \pi_i(t) \log \left(\frac{\pi_i(t)}{\pi^{\rm st}_i}\right)\nonumber\\
&=& \sum_{i,j}  (\pi_i(t) k_{i \to j} - \pi_j(t) k_{j \to i})  \log \left(\frac{\pi_i(t)}{\pi^{\rm st}_i}\right)\nonumber\\
&=& -\sum_{i,j}  \pi_j(t) k_{j \to i}   \log \left(\frac{\pi_i(t) \pi^{\rm st}_j}{\pi_j(t) \pi^{\rm st}_i}\right).
\end{eqnarray}
Now we use the fact that
\begin{equation}
-\ln x \ge 1-x  \label{log}
\end{equation}
to obtain
\begin{eqnarray}
\dot S_{\rm cyc} &\ge& \sum_{i,j}  \pi_j(t) k_{j \to i}   \left[ 1 -   \frac{\pi_i(t) \pi^{\rm st}_j}{\pi_j(t) \pi^{\rm st}_i}  \right] \times {\rm const.}\nonumber\\
&=& \left[ \sum_{i,j}  \pi_j(t) k_{j \to i}  -  \sum_{i} \frac{\pi_i(t) }{ \pi^{\rm st}_i}  \sum_{j} \pi^{\rm st}_j k_{j \to i} \right] \times {\rm const.}\nonumber\\
&=& \sum_{i} \dot \pi_i(t)   \times {\rm const.}=0.
\end{eqnarray} 

Since $S_{\rm cyc}$ has the form of a negative relative entropy, which measures the similarity of the distribution $\pi_i(t)$ to $\pi^{\rm st}_i$, the result here tells us that $\pi$ becomes more similar to $\pi^{\rm st}_i$ as time proceeds. Note that only the balance condition for $\pi^{\rm st}_i$ was used in the derivation. Therefore, the result holds even for the case where $\pi_i(t)$ does not converge to $\pi^{\rm st}_i$. The result here can be considered a consequence of a detailed fluctuation theorem~\cite{episto}. 
\section{Proof of $\Sigma \ge S_{\rm cyc}/\Delta t$ or $\Sigma \ge \dot S_{\rm cyc}$ }\label{sigcyc}
Here we show that the Schnakenberg entropy production $\Sigma$, defined by Eq.(\ref{sch}) and Eq.(\ref{schd}) for continuous-time and discrete-time Markov models, respectively, is not less than the increase of $S_{\rm cyc}$ with time. Regarding discrete time, by using the definition of $\Sigma$ in Eq.(\ref{schd}) and the expression for $\Delta S_{\rm cyc}/{\Delta t}$ in Eq.(\ref{dScy}),  we have
\begin{eqnarray}
\Sigma - \frac{\Delta S_{\rm cyc}}{\Delta t} &=& \sum_{i,j} \pi_i(t) k_{i \to j} \log \left(\frac{ \pi_i(t) k_{i \to j}}{\pi_j(t + \Delta t)  k_{j \to i}} \right) - \sum_{i,j} \pi_i(t) k_{i \to j} \log \left(\frac{ \pi_i(t)\pi^{\rm st}_j}{\pi_j(t + \Delta t)  \pi^{\rm st}_i} \right) \nonumber\\
&=& - \sum_{i,j} \pi_i(t) k_{i \to j} \log \left(\frac{ \pi^{\rm st}_j k_{j \to i}}{\pi^{\rm st}_i  k_{i \to j}} \right). \label{entpro}
\end{eqnarray}
Note that $\Delta t$ disappears in Eq.(\ref{entpro}). Therefore, the expression for $\Sigma - \dot S_{\rm cyc}$ in the continuum model is also given by Eq.(\ref{entpro}). Using Eq.(\ref{log}), we obtain
\begin{eqnarray}
\Sigma - \frac{\Delta S_{\rm cyc}}{\Delta t} &\ge&
\sum_{i,j} \left[ \pi_i(t) k_{i \to j}  -  \frac{\pi_i(t) \pi^{\rm st}_j k_{j \to i}}{\pi^{\rm st}_i}  \right] \times {\rm constant}\nonumber\\
&=& \sum_j \frac{\Delta \pi_j(t)}{\Delta t} = 0.  
\end{eqnarray}
This also shows that $\Sigma - \dot S_{\rm cyc} \ge 0$ in a continuous-time model\cite{hg}. Therefore, combining the result with that in Appendix \ref{entcyc}, we have proved that
\begin{equation}
\Sigma \ge \frac{S_{\rm cyc}}{\Delta t} \ge 0
\end{equation}
for a discrete-time model, and
\begin{equation}
\Sigma \ge \dot S_{\rm cyc} \ge 0
\end{equation}
for a continuous-time model. These results can be considered a consequence of a detailed fluctuation theorem\cite{episto}. The explicit connections between the production of entropy components defined in this work, and those in ref.\cite{episto}, are given in appendix \ref{adna}. 
\section{Proof of $\Delta S_{\rm hid}/\Delta t \ge 0$ or $\dot S_{\rm hid} \ge 0$ in the cyclic regime}\label{dhid}
We prove that the hidden entropy production is nonnegative in the cyclic regime\cite{kaw,zmh,epi,chun}. An important contribution of the current work is that the hidden entropy production can be explicitly expressed as $\dot S_{\rm hid}$ with $S_{\rm hid}$ having an analytic expression constructed from the probability distributions of the extended model. Regarding the extended model with discrete time, we construct the quantity
\begin{equation}
\tilde \Sigma \equiv \Delta t^{-1}\sum_{i,j} \pi_i(t) q(i \to j; t) \log \left(\frac{ \pi_i(t) q(i \to j;t)}{\pi_j(t+\Delta t)  q(j \to i;t+\Delta t)} \right),
\label{sch2}
\end{equation}
where $q(i \to j; t) \equiv \delta_{i,j} + w(i \to j;t) \Delta t$ is the time-dependent transition probability of the reduced model, with $w(i \to j;t)$ being the time-dependent transition rate. Note that $\tilde \Sigma$ is reduced to $\Sigma$ in the cyclic regime where $w(i \to j;t) \simeq k_{i \to j}$. 

Now, we define a two-variable function $\phi(x,y) \equiv x \ln (x /y) $. This function is convex in the range $0\le x, y \le 1$, implying that
\begin{equation}
\phi (\sum_\beta a_\beta x_\beta,\sum_\gamma a_\gamma y_\gamma) \le \sum_\beta a_\beta \phi (x_\beta,y_\beta)   \label{cnv2d}
\end{equation}
for any values of $0 \le x_\beta,y_\beta \le 1$ and $0 \le a_\beta \le 1$ with $\sum_\beta a_\beta = 1$. The convexity of $\phi(x,y)$ can be shown by utilizing the fact that a multivariate function is convex if and only if its Hessian matrix is positive semidefinite. In fact, we find that the Hessian matrix $H(x,y)$ is
\begin{equation}
H(x,y)\equiv \begin{pmatrix} 
\frac{\partial^2 \phi(x,y)}{\partial x^2}  & \frac{\partial^2 \phi(x,y)}{\partial x \partial y}  \\
\frac{\partial^2 \phi(x,y)}{\partial y \partial x} & \frac{\partial^2 \phi(x,y)}{\partial y^2} 
\end{pmatrix} = \begin{pmatrix} 
\frac{1}{x}  & -\frac{1}{y}  \\
-\frac{1}{y} & \frac{x}{ y^2} 
\end{pmatrix}, 
\end{equation}
whose eigenvalues are $\lambda_1=0$ and $\lambda_2 = x/y^2+1/x$, both of which are nonnegative for $0 \le x,y \le 1$. 

For a given pair of indices $i$ and $j$ of the reduced Markov model, we now use $\beta = ({\bf X},{\bf Y})$, $x_\beta = \Pi_{(i,{\bf X})}(t) P_{(i,{\bf X} \to j,{\bf Y})}$, $y_\beta = \Pi_{(j,{\bf Y})}(t + \Delta t) P_{(j,{\bf Y} \to i,{\bf X})}$ with $P_{(i,{\bf X} \to j,{\bf Y})} \equiv  W_{(i,{\bf X} \to j,{\bf Y})}\Delta t + \delta_{i,j} \delta_{\bf X,\bf Y}$, and $a_\beta = 1/(n_i n_j)$ with $n_i$ denoting the total number of ${\bf X}$ values corresponding to $i$. The left-hand side of Eq.(\ref{cnv2d}) then becomes
\begin{eqnarray}
\phi (\sum_\beta a_\beta x_\beta,\sum_\gamma a_\gamma y_\gamma) &=&
\phi (\sum_{{\bf X},{\bf Y}} \Pi_{(i,{\bf X})}(t) P_{(i,{\bf X} \to j,{\bf Y})}/n_i n_j ,\sum_{{\bf \tilde X},{\bf \tilde Y}} \Pi_{(j,{\bf \tilde Y})}(t+ \Delta t)P_{(j,{\bf \tilde Y} \to i,{\bf \tilde X})}/n_i n_j) \nonumber\\
&=& \phi (\pi_i (t) q (i  \to j; t)/n_i n_j , \pi_j (t+ \Delta t) q (j  \to i; t+ \Delta t)/n_i n_j) \nonumber\\
&=& \frac{1}{n_i n_j} \pi_i (t) q (i  \to j; t) \ln \frac{\pi_i (t) q (i  \to j; t)}{\pi_j (t + \Delta t) q (j  \to i; t + \Delta t)}
,
\end{eqnarray}
and the right-hand side of Eq.(\ref{cnv2d}) becomes
\begin{eqnarray}
\sum_\beta a_\beta \phi (x_\beta,y_\beta) &=&
\sum_{{\bf X},{\bf Y}} \frac{1}{n_i n_j } \phi ( \Pi_{(i,{\bf X})}(t) P_{(i,{\bf X} \to j,{\bf Y})}, \Pi_{(j,{\bf Y})}(t+ \Delta t) P_{(j,{\bf Y} \to i,{\bf X})})  \nonumber\\
&=& \frac{1}{n_i n_j} \sum_{{\bf X},{\bf Y}}  \Pi_{(i,{\bf X})}(t) P_{(i,{\bf X} \to j,{\bf Y})}  \ln \frac{\Pi_{(i,{\bf X})}(t) P_{(i,{\bf X} \to j,{\bf Y})}}{\Pi_{(j,{\bf Y})}(t+ \Delta t) P_{(j,{\bf Y} \to i,{\bf X})}}.
\end{eqnarray}
Therefore, using the inequality Eq.(\ref{cnv2d}) and summing over $i$ and $j$, we get
\begin{equation}
\tilde \Sigma  \le \frac{\Delta S_{\rm tot}}{\Delta t}. \label{hdprd}
\end{equation}
The continuous counterpart of Eq.(\ref{hdprd}),
\begin{equation}
\tilde \Sigma  \equiv \sum_{i,j} \pi_i(t) w(i \to j; t) \log \left(\frac{ \pi_i(t) w(i \to j;t)}{\pi_j(t)  w(j \to i;t)} \right) \le \dot S_{\rm tot}, \label{hdprc}
\end{equation}
is obtained straightforwardly by taking the limit $\Delta t \to 0$ of Eq.(\ref{hdprd}), which has also been derived directly~\cite{zmh}. The discrete generalization Eq.(\ref{hdprd}) is a new contribution of the current work.
 
In the cyclic regime, we have
\begin{equation}
\frac{\Delta S_{\rm hid}}{\Delta t}  = \frac{\Delta S_{\rm tot}}{\Delta t} - \Sigma_{\rm exact} \simeq \frac{\Delta S_{\rm tot}}{\Delta t} - \Sigma \simeq \frac{\Delta S_{\rm tot}}{\Delta t} - \tilde \Sigma \ge 0 
\end{equation}
for the discrete-time model and 
\begin{equation}
\dot S_{\rm hid} = \dot S_{\rm tot} - \Sigma_{\rm exact} \simeq \dot S_{\rm tot} - \Sigma \simeq \dot S_{\rm tot} - \tilde \Sigma \ge 0
\end{equation}
for the continuous-time model, where the first approximation in each of these equations is valid if Eq.(\ref{condom2}) is satisfied.
 
Note that Eq.(\ref{hdprd}) or Eq.(\ref{hdprc}) is an exact result that hold without the conditions Eq.(\ref{simp2}) and Eq.(\ref{condom2}), which is a consequence of an integral fluctuation theorem~\cite{kaw}. This leads to an alternative definition of the hidden entropy, $\tilde S_{\rm hid}(t) \equiv S_{\rm tot} - \int^t_0 \tilde \Sigma(t') dt'$ (continuous time) or $\tilde S_{\rm hid} \equiv S_{\rm tot} - \Delta t  \sum_{j=0}^n \tilde \Sigma(j \Delta t)$ (discrete time). We then have ${\dot {\tilde S}}_{\rm hid} = \dot S_{\rm tot} - \tilde \Sigma$ (continuous time) or ${\Delta {\tilde S}}_{\rm hid}/\Delta t = \Delta S_{\rm tot}/\Delta t - \tilde \Sigma$ (discrete time). However, it does not seem straightforward to find an analytic expression for  $\tilde S_{\rm hid}(t)$.

\section{Derivation of ${\Delta S_{\rm tot}}/{\Delta t} - {\Delta S_{\rm hid}}/{\Delta t} = \Sigma_{\rm exact}$ (Eq.(\ref{fin1})) or $\dot S_{\rm tot} - \dot S_{\rm hid} = \Sigma_{\rm exact}$ (Eq.(\ref{entpr}))}\label{totsig}
By applying Eq.(\ref{entpd}) to the extended model, we have
\begin{eqnarray}
\frac{\Delta S_{\rm tot}}{\Delta t}
&=&   \sum_{i,j,{\bf X},{\bf Y}}  \Pi_{(i,{\bf X})}(t) W_{(i,{\bf X}) \to (j,{\bf Y})} \log \frac{\Pi_{(i,{\bf X})}(t ) W_{(i,{\bf X}) \to (j,{\bf Y})}}{\Pi_{(j,{\bf Y})}(t+ \Delta t) W_{(j,{\bf Y}) \to (i,{\bf X})}}\nonumber\\
&+& \Delta t^{-1} \sum_{i,{\bf X} } \Pi_{(i,{\bf X})}(t)  \log \frac{\Pi_{i,{\bf X}}(t) }{\Pi_{i,{\bf X}}(t+ \Delta t)}\label{totd}
\end{eqnarray} 
for the discrete-time Markov model. By performing the decomposition $\Pi_{(i,{\bf X})}(t) \equiv \pi_i (t) \Pi_{({\bf X}/i)}(t)$, Eq.(\ref{totd}) is rewritten as
\begin{eqnarray}
\frac{\Delta S_{\rm tot}}{\Delta t}
&=&  \sum_{i,j,{\bf X}}  \Pi_{(i,{\bf X})}(t) W_{(i,{\bf X}) \to (j,{\bf Y})} \log \frac{\pi_{i}(t) W_{(i,{\bf X}) \to (j,{\bf Y})}}{\pi_{j}(t+ \Delta t) W_{(j,{\bf Y}) \to (i,{\bf X})}} \nonumber\\
&&+ \sum_{i,j,{\bf X},{\bf Y}} \Pi_{(i,{\bf X})}(t) W_{(i,{\bf X}) \to (j,{\bf Y})}  \log \frac{\Pi_{({\bf X}/i)}(t)}{\Pi_{({\bf Y}/j)}(t+ \Delta t)} \nonumber\\
&&+ \Delta t^{-1} \sum_{i } \pi_{i}(t)  \log \frac{\pi_{i}(t) }{\pi_{i}(t+ \Delta t)}
\nonumber\\
&&+ \Delta t^{-1}\sum_{i,{\bf X} } \Pi_{(i,{\bf X})}(t)  \log \frac{\Pi_{{\bf X}/i}(t) }{\Pi_{{\bf X}/i}(t+ \Delta t)}\nonumber\\
&=&  \Sigma_{\rm exact} +  \sum_{i,j,{\bf X},{\bf Y}} \Pi_{(i,{\bf X})}(t) W_{(i,{\bf X}) \to (j,{\bf Y})} \log \frac{\Pi_{({\bf X}/i)}(t) }{\Pi_{({\bf Y}/j)}(t+ \Delta t)}\nonumber\\
&&+ \Delta t^{-1}\sum_{i,{\bf X} } \Pi_{(i,{\bf X})}(t)  \log \frac{\Pi_{{\bf X}/i}(t) }{\Pi_{{\bf X}/i}(t+ \Delta t)}\nonumber\\
&=&  \Sigma_{\rm exact} -  \sum_{i,j,{\bf X},{\bf Y}} \Pi_{(i,{\bf X})}(t) W_{(i,{\bf X}) \to (j,{\bf Y})} \log \Pi_{({\bf Y}/j)}(t+ \Delta t)\nonumber\\
&&+ \Delta t^{-1}\sum_{i,{\bf X} } \Pi_{(i,{\bf X})}(t)  \log \frac{\Pi_{{\bf X}/i}(t) }{\Pi_{{\bf X}/i}(t+ \Delta t)},
\label{apph1}
\end{eqnarray} 
where we used  $\sum_{j,{\bf Y}} W_{(i,{\bf X}) \to (j,{\bf Y})}=0$ to derive the last line, and $\Sigma_{\rm exact}$ for the discrete-time model is defined as
\begin{equation}
\Sigma_{\rm exact} \equiv \sum_{i,j,{\bf X}}  \Pi_{(i,{\bf X})}(t) W_{(i,{\bf X}) \to (j,{\bf Y})} \log \frac{\pi_{i}(t) W_{(i,{\bf X}) \to (j,{\bf Y})}}{\pi_{j}(t+ \Delta t) W_{(j,{\bf Y}) \to (i,{\bf X})}} + \Delta t^{-1} \sum_{i } \pi_{i}(t)  \log \frac{\pi_{i}(t) }{\pi_{i}(t+ \Delta t)}.
\end{equation}
Now, using the definition of the hidden entropy,
\begin{equation}
S_{\rm hid} \equiv  -\sum_{i,{\bf X}} \Pi_{(i,{\bf X})}(t)  \log \Pi_{({\bf X}/i)}(t),\label{hid2}
\end{equation}
we obtain
\begin{eqnarray}
\Delta S_{\rm hid}
&=&  -\sum_{i,{\bf X}}  \Pi_{(i,{\bf X})}(t + \Delta t)  \log \Pi_{({\bf X}/i)}(t + \Delta t) + \sum_{i,{\bf X}} \Pi_{(i,{\bf X})}(t)  \log \Pi_{({\bf X}/i)}(t)\nonumber\\
&=& -\sum_{i,{\bf X}}  \Pi_{(i,{\bf X})}(t + \Delta t)  \log \Pi_{({\bf X}/i)}(t + \Delta t) +\sum_{i,{\bf X}}  \Pi_{(i,{\bf X})}(t )  \log \Pi_{({\bf X}/i)}(t + \Delta t)\nonumber\\
&& - \sum_{i,{\bf X}} \Pi_{(i,{\bf X})}(t)  \log \Pi_{({\bf X}/i)}(t + \Delta t) + \sum_{i,{\bf X}} \Pi_{(i,{\bf X})}(t)  \log \Pi_{({\bf X}/i)}(t)\nonumber\\
&=& -\sum_{i,{\bf X}}  
\Delta \Pi_{(i,{\bf X})}(t)  \log \Pi_{({\bf X}/i)}(t + \Delta t) \nonumber\\
&&  + \sum_{i,{\bf X}} \Pi_{(i,{\bf X})}(t)  \log \frac{\Pi_{({\bf X}/i)}(t)}{\Pi_{({\bf X}/i)}(t + \Delta t)}\nonumber\\
&=& - \Delta t \sum_{i,j,{\bf X},{\bf Y}}  \Pi_{(j,{\bf Y})}(t) W_{(j,{\bf Y}) \to (i,{\bf X})}  \log \Pi_{({\bf X}/i)}(t + \Delta t)\nonumber\\
&&+ \sum_{i,{\bf X}} \Pi_{(i,{\bf X})}(t)  \log \frac{\Pi_{({\bf X}/i)}(t)}{\Pi_{({\bf X}/i)}(t + \Delta t)}.
\label{apph2}
\end{eqnarray}
From Eq.(\ref{apph1}) and Eq.(\ref{apph2}), we see that
\begin{equation}
\frac{\Delta S_{\rm tot}}{\Delta t} = \Sigma_{\rm exact} + \frac{\Delta S_{\rm hid}}{\Delta t}.\label{stot2}
\end{equation}
for the discrete-time model.

The result for the continuous time model,
\begin{equation}
\dot S_{\rm tot} = \Sigma_{\rm exact} + \dot S_{\rm hid},
\end{equation}
is obtained by taking the limit $\Delta t \to 0$ in Eq.(\ref{stot2}), where we now have
\begin{equation}
\Sigma_{\rm exact} \equiv  \sum_{i,j,{\bf X},{\bf Y}}   \Pi_{(i,{\bf X})}(t) W_{(i,{\bf X}) \to (j,{\bf Y})}  \log \frac{\pi_i (t) W_{(i,{\bf X}) \to (j,{\bf Y})} }{\pi_j (t) W_{(j,{\bf Y}) \to (i,{\bf X})} },
\end{equation}
for the continuous-time model.

\section{Derivation of expressions for $\Delta S_{\rm cyc}/\Delta t$ and $\Delta S_{\rm closed}/\Delta t$ (Eq.(\ref{entpd}))}\label{rem}
We have
\begin{eqnarray}
\frac{\Delta S_{\rm closed}}{\Delta t} &\equiv& \left[ S_{\rm closed}(t+\Delta t) - S_{\rm closed}(t)\right]\frac{1}{\Delta t} \nonumber\\
&=& -\sum_i \frac{\pi_i(t+\Delta t)}{\Delta t} \log \left(\frac{\pi_i(t+ \Delta t)}{\pi^{\rm eq}_i}\right) + \sum_i \frac{\pi_i(t)}{\Delta t} \log \left(\frac{\pi_i(t)}{\pi^{\rm eq}_i}\right) \nonumber\\
&=& -\sum_i \frac{\pi_i(t+\Delta t)}{\Delta t} \log \left(\frac{\pi_i(t+ \Delta t)}{\pi^{\rm eq}_i}\right) + \sum_i \frac{\pi_i(t+\Delta t)}{\Delta t} \log \left(\frac{\pi_i(t)}{\pi^{\rm eq}_i}\right) \nonumber\\
&& -\sum_i \frac{\pi_i(t+\Delta t)}{\Delta t} \log \left(\frac{\pi_i(t)}{\pi^{\rm eq}_i}\right)  + \sum_i \frac{\pi_i(t)}{\Delta t} \log \left(\frac{\pi_i(t)}{\pi^{\rm eq}_i}\right) \nonumber\\
&=& {\Delta t}^{-1}\sum_i {\pi_i(t+\Delta t)} \log \left(\frac{\pi_i(t)}{\pi^i (t+\Delta t)}\right)\nonumber\\
&&-\sum_i \frac{\Delta \pi_i(t)}{\Delta t} \log \left(\frac{\pi_i(t)}{\pi^{\rm eq}_i}\right) \nonumber\\ 
&=& -\sum_{i,j}  \pi_j(t) k_{j \to i} \log \left(\frac{\pi_i(t)}{\pi^{\rm eq}_i}\right) + {\Delta t}^{-1} \sum_i \pi_i(t+\Delta t)\log \left(\frac{\pi_i(t)}{\pi^i (t+\Delta t)}\right)\nonumber\\
&=& -{\Delta t}^{-1} \sum_{i,j} ( \pi_j(t) p_{j \to i} - \pi_i(t) p_{i \to j})  \log \left(\frac{\pi_i(t)}{\pi^{\rm eq}_i}\right) \nonumber\\
&&+ {\Delta t}^{-1} \sum_{i,j}  \pi_j(t) p_{j \to i} \log \left(\frac{\pi_i(t)}{\pi^i (t+\Delta t)}\right)\nonumber\\
&=& -{\Delta t}^{-1} \sum_{i,j} \pi_j(t) p_{j \to i}   \log \left(\frac{\pi_i(t) \pi^{\rm eq}_j}{\pi_j(t) \pi^{\rm eq}_i}\right) - {\Delta t}^{-1} \sum_{i,j}  \pi_j(t) p_{j \to i} \log \left(\frac{\pi_i(t+\Delta t)}{\pi^i (t)}\right)\nonumber\\
&=& -{\Delta t}^{-1} \sum_{i,j} \pi_j(t) p_{j \to i}   \log \left(\frac{\pi_i(t +\Delta t) \pi^{\rm eq}_j}{\pi_j(t) \pi^{\rm eq}_i}\right). \label{dScl}
\end{eqnarray}
By using detailed balance condition, we can rewrite Eq.(\ref{dScl}) as
\begin{eqnarray}
\frac{\Delta S_{\rm closed}}{\Delta t} &=& {\Delta t}^{-1} \sum_{i,j} \pi_j(t) p_{j \to i}   \log \left(\frac{\pi_j(t) p_{j \to i}}{\pi_i(t +\Delta t) p_{i \to j}}\right) \nonumber\\
&=& \sum_{i,j} \pi_j(t) k_{j \to i}   \log \left(\frac{\pi_j(t) k_{j \to i}}{\pi_i(t +\Delta t) k_{i \to j}}\right) + {\Delta t}^{-1} \sum_{i} \pi_i(t)  \log \left(\frac{\pi_i(t) }{\pi_i(t +\Delta t) }\right)
.
\end{eqnarray}
Note that Eq.(\ref{dScl}) is valid even for the model without detailed balance, with the change of notation $S_{\rm closed} \to S_{\rm cyc}$  and $\pi^{\rm eq}_i \to \pi^{\rm st}_i $:
\begin{eqnarray}
\frac{\Delta S_{\rm cyc}}{\Delta t} &=& -{\Delta t}^{-1} \sum_{i,j} \pi_j(t) p_{j \to i}   \log \left(\frac{\pi_i(t +\Delta t) \pi^{\rm st}_j}{\pi_j(t) \pi^{\rm st}_i}\right)\nonumber\\
&=& - \sum_{i,j} \pi_j(t) k_{j \to i}   \log \left(\frac{\pi_i(t +\Delta t) \pi^{\rm st}_j}{\pi_j(t) \pi^{\rm st}_i}\right) \nonumber\\
&& -{\Delta t}^{-1} \sum_{i} \pi_i(t)  \log \left(\frac{\pi_i(t +\Delta t) }{\pi_i(t) }\right). \label{dScy}
\end{eqnarray}

\section{Coarse-graining of Shannon entropy}\label{shanlee}
We assume that $c$ labels a macrostate, which is an aggregate of microstates, and that an additional index $\alpha=1,\cdots \Omega_c$ is required to completely specify a microstate. Then the Shannon entropy of the closed system is
\begin{equation}
S_{\rm shan} = - \sum_c \sum_{\alpha=1}^{\Omega_c} p(c,\alpha)(t) \log p(c,\alpha)(t). \label{shanshan}
\end{equation}
where $p(c,\alpha)(t)$ is the probability that the system is at the microstate $(c,\alpha)$ at time $t$. 

Now we assume that the microstates for given $c$ are at local equilibrium, meaning that $\sum_{\alpha=1}^{\Omega_c} p(c,\alpha) \log p(c,\alpha)$ is maximized for each $c$. Thus, $c$ is a slow variable and $\alpha$ is a fast variable. We first make decomposition $p(c,\alpha) = P_c p_{\alpha/c}$ where $P_c \equiv \sum_\alpha p(c,\alpha)$ and $p_{\alpha/c} \equiv  p(c,\alpha)/P_c$ are the marginal and the conditional probabilities, respectively. Then $S_{\rm shan}$ is rewritten as~\cite{lee}
\begin{equation}
S_{\rm shan} = - \sum_{c} P_c \log P_c - \sum_c P_c \sum_{\alpha=1}^{\Omega_c }p_{\alpha/c} \log p_{\alpha/c}. \label{shan2}
\end{equation}
To obtain local equilibrium, Eq.(\ref{shan2}) is maximized with respect to $p_{c/\alpha}$,  under the normalization constraint
\begin{equation}
\sum_{\alpha=1}^{\Omega_c} p_{\alpha/c} = 1, \label{norm2}
\end{equation}
treating $P_c$ as a constant.

Introducing the Lagrange multiplier $\lambda_c$ for the constraint Eq.(\ref{norm2}), we take the derivative of the target function\footnote{We assume that the summation range of $(c,\alpha)$ in Eq.(\ref{shanshan}) goes only over the states that satisfy appropriate constraints, such as those on the values of conserved quantities, so that the Lagrange multipliers for these additional constraints do not have to be introduced explicitly.}
\begin{equation}
F(p_{\alpha/c}) \equiv - \sum_{c} P_c \log P_c - \sum_c P_c \sum_{\alpha=1}^{\Omega_c } p_{\alpha/c} \log p_{\alpha/c} + \sum_c \lambda_c \sum_{\alpha=1}^{\Omega_c} p_{\alpha/c},
\end{equation}
with respect to $p_{\alpha/c}$ and set it to zero~\cite{lee}:
\begin{equation}
\frac{\partial F}{\partial p_{\alpha/c}}= -  P_c (1 + \ln p_{\alpha/c})(\ln B)^{-1} +  \lambda_c = 0,
\end{equation}
leading to
\begin{equation}
 p_{\alpha/c} =  \exp(\lambda_c \ln B /P_c -1),\label{sol1}
\end{equation}
where $B$ is the base of the log function.
The second derivative of $F(p_{\alpha/c})$ is
\begin{equation}
\frac{\partial^2 F}{\partial p_{\alpha/c} \partial p_{\beta/c}}= -\frac{P_c}{p_{\alpha/c}} (\ln B)^{-1} \delta_{\alpha,\beta} \le 0,
\end{equation}
implying that the solution Eq.(\ref{sol1}) is the maximum rather than the minimum or extremum. The Lagrange multiplier $\lambda_c$ is obtained by the normalization constraint Eq.(\ref{norm2}), so that
\begin{equation}
p_{\alpha/c} = \Omega_c^{-1}. \label{pom}
\end{equation}
That is, all the microstates for given macrostate state $c$ are occupied with equal probability. Substituting Eq.(\ref{pom}) into Eq.(\ref{shan2}), we get~\cite{lee}
\begin{equation}
S_{\rm closed} = - \sum_{c} P_c \log P_c + \sum_c P_c S_c + {\rm constant}, \label{closd}
\end{equation}
where $S_{\rm closed}$ is the entropy of the closed system at local equilibrium. Note that now the fast variable $\alpha$ does not explicitly appear in $S_{\rm closed}$. The equilibrium distribution over the slow variable $c$ is now obtained by maximizing $S_{\rm closed}$ with respect to $P_c$, under the normalization constraint $\sum_c P_c = 1$, which is
\begin{equation}
P^{\rm eq}_c \propto \Omega_c. \label{gbol}
\end{equation}
The equilibrium distribution $P^{\rm eq}_c$ reduces to the Boltzmann distribution in the special case when number of fast degrees of freedom per energy is effectively infinite~(See Appendix \ref{open}). 
 From Eq.(\ref{pom}) and Eq.(\ref{gbol}), we also get
 \begin{equation}
p^{\rm eq}(c, \alpha) \propto \Omega_c \Omega_c^{-1} = {\rm constant}, 
\end{equation}
the well-known result that at the equilibrium, all the microstates of the closed system consistent with given constraints are occupied with equal probabilities~\cite{lee,tolman}. 
\section{Open system}\label{open}
 The Markov model of an open system, often discussed in the literature~\cite{seif1,seif2,hg,gq,hs,sp}, constitutes a special case of the Markov model presented in the current work, whose details are given below.  First, it is assumed that the total energy $E_{\rm closed}$ of the closed system is additive:
\begin{equation}
E^{\rm closed}_{c,\alpha} = E^{\rm slow}_c + E^{\rm fast}_\alpha,
\end{equation}
where, following the notation of Appendix \ref{shanlee}, $c$ and $\alpha$ denote the slow and fast variables, respectively. 
Second, it is assumed that for given values of $E^{\rm slow}_c=E_1$ and $E^{\rm fast}_\alpha=E_2$, the total number of microstates $g(E_1,E_2)$ is multiplicative:
\begin{equation}
g(E_1, E_2) = g_{\rm slow}(E_1) \times g_{\rm fast}(E_2), 
\end{equation}
where $g_{\rm slow}(E_1)$ $(g_{\rm fast}(E_2))$ is the number of states with energy $E_1$ ($E_2$) that the slow (fast) variable can take. 
Finally, it is assumed that the number of fast degree of freedom is effectively infinite for a given energy value so that its temperature,
\begin{equation}
T \equiv \left(\frac{d} {d E_2} \log g_{\rm fast}(E_2)\right)^{-1}
\end{equation}
can be regarded as a constant. In this case, we call the fast variables as a heat bath of temperature $T$, and we call the slow variables an open system in contact with the heat bath. The equilibrium distribution in Eq.(\ref{gbol}) now reduces to the Boltzmann distribution:
\begin{equation}
P^{\rm eq}_c \propto \Omega_c =g( E_{\rm close}-E_c) \simeq B^{-E_c/T} \times {\rm constant}, \label{fact}
\end{equation} 
where $B$ is the base of the log function.
 Using Eq.(\ref{fact}), the total entropy of the closed system in Eq.(\ref{sysmed}) is now rewritten as
\begin{equation}
S_{\rm closed} = -F/T + {\rm constant}, \label{free2}
\end{equation}
where 
\begin{equation}
F \equiv \sum_c \pi_c E_c - T  \sum_c \pi_c(t) \log \pi_c(t), \label{free3} 
\end{equation} 
is a free energy that is well defined even out of equilibrium~\cite{gq,ne}. Therefore, the entropy production can also be called the free energy reduction. 

After decomposing the slow degrees of freedom into those of the driver and the driven system, $c=(i,{\bf X})$, Eq.(\ref{free3}) is rewritten as
\begin{equation}
F = -T S_{\rm bol}-T S_{\rm shan}-T S_{\rm hid}, \label{dcmp4} 
\end{equation} 
where the definitions of $S_{\rm shan}$ and $S_{\rm hid}$ are unchanged from those in the general case, Eq.(\ref{shan1}) and Eq.(\ref{hid1}), but $S_{\rm bol}$ is now rewritten as
\begin{equation}
S_{\rm bol} = \sum_{(i,{\bf X})} \Pi_{(i,{\bf X})} \log \Omega_{(i,{\bf X})}  \simeq - \sum_{(i,{\bf X})} \Pi_{(i,{\bf X})} E_{(i,{\bf X})}/T +{\rm constant}.
\end{equation}
As in the general case, the free energy excluding the contribution from the driver degrees of freedom,  $\tilde F \equiv F + T S_{\rm hid}$, is often considered. From the general result, we already know that
\begin{equation}
T \Sigma \simeq -\dot {\tilde F} =  T \dot S_{\rm shan} + T \dot S_{\rm bol} 
\end{equation}
in the cyclic regime where the conditions Eq.(\ref{simp2}) and Eq.(\ref{condom2}) hold. Because fast and slow variables are considered to represent the system and the bath, respectively, $\dot S_{\rm bol}$ gives the production of entropy in the bath. Therefore, 
\begin{equation}
q_{\rm tot} \equiv T \dot S_{\rm bol} \simeq T (\Sigma - \dot S_{\rm shan}), \label{theh}
\end{equation}
is considered as the heat production in the bath~\cite{seif1,seif2,hg,gq,hs,sp}.

From the alternative decomposition $S_{\rm shan} + S_{\rm bol} = S_{\rm cyc} + S_{\rm hk}$, we see that when the system reaches the steady state $\pi_i(t) \simeq \pi^{\rm st}_i$, we have $\dot S_{\rm shan} \simeq \dot S_{\rm cyc} \simeq 0$, so that $q_{\rm tot}= T \dot S_{\rm bol} \simeq T \dot S_{\rm hk}$. Therefore,  we see that the housekeeping heat production, defined as
\begin{equation}
q_{\rm hk} \equiv T \dot S_{\rm hk}, 
\end{equation}
is the rate of heat being dissipated while the quasi-steady state is maintained. The excess heat production $q_{\rm ex}$ is the rate of extra heat dissipation during the approach to the quasi-steady state:
\begin{eqnarray}
q_{\rm ex} &\equiv& q_{\rm tot} - q_{\rm hk} = T (\dot S_{\rm bol} - \dot S_{\rm hk})\nonumber\\
&=& T (\dot S_{\rm cyc} - \dot S_{\rm shan})\nonumber\\
&=& T \frac{d}{dt}\sum_i \pi_i(t) \log \pi^{\rm st}_i \nonumber\\
&=& T \sum_{i,j} \pi_j(t) k_{j \to i} \log \pi^{\rm st}_i. \label{qex}
\end{eqnarray}
\section{The Langevin and Fokker-Planck equations}\label{lang}
 Overdamped Langevin dynamics on a circle, described by the equation
\begin{equation}
\gamma \dot x = -\frac{\partial U(x,\lambda)}{\partial x} + f + \zeta(t)\label{lg}
\end{equation}
where $\lambda$ is a control parameter, $f$ is a non-conservative force, and  $\zeta(t)$ is a Gaussian white noise satisfying the relation
\begin{equation}
\langle \zeta(t) \zeta(t')\rangle = 2 \gamma T,
\end{equation}
has often been considered in the literature~\cite{seif1,seif2,hs,sp,hy2}. 
The time evolution of the probability density $\rho(x)$ of this system is described by the  Fokker-Planck equation:
\begin{equation}
\dot \rho = -\partial_x (F(x) \rho) +  B \partial_x^2 \rho, \label{fp}
\end{equation} 
where $F(x) \equiv -\partial_x U/\gamma+f/\gamma$ and $B \equiv T/ \gamma$~\cite{zwan}. 

The Fokker-Planck equation  Eq.(\ref{fp}) is nothing but a Markov process with a continuous state index $x$, and can be easily obtained from the Markov model with discrete state indices by taking an  appropriate continuum limit, where the fast degree of freedom is considered to form a heat bath of constant temperature T~(Appendix \ref{open}).  More concretely, we identify the state $i=0,1, \cdots N-1$ with a position on a circle, with equal spacing $\ell$  between the neighboring sites, which will shrink to zero in the continuum limit. We also initially consider discrete time with a time step $\Delta t$, so that we can take a simultaneous limit of $\ell \to 0$ and $\Delta t \to 0$, with $\ell \propto \Delta t^\alpha$. The dynamics of the system is described by Eq.(\ref{mard}), with the condition that $k_{i \to j}=0$ unless $j=\left[(i\pm 1) \mod N\right]$:
\begin{equation}
\frac{\Delta \pi_i(t)}{\Delta t} = \sum_{|j-i| \leq 1} \pi_j(t ) k_{j \to i} = \pi_{i-1}(t ) k_{(i-1) \to i} +\pi_{i+1}(t ) k_{(i+1) \to i} -  \pi_{i}(t)(k_{i \to (i-1)} + k_{i \to (i+1)} ). \label{rw1}  
\end{equation}
Now we define the symmetric and antisymmetric components of $k_{i \to j}$ as
\begin{eqnarray}
s_i &\equiv (k_{i \to (i+1)} + k_{i \to (i-1)})/2 \nonumber\\
a_i &\equiv (k_{i \to (i+1)} - k_{i \to (i-1)})/2,
\end{eqnarray}
so that $k_{i \to j} = s_i \pm a_i$ for $j= i \pm 1$. 
The antisymmetric component $a_i$ describes the drift of the particle in one direction, which will be shown to be proportional to the $f-\partial_x U$ term in the Langevin dynamics,  and the symmetric component $s_i$ describes the diffusion due to random fluctuations $\zeta$. Considering the fact that the strength of $\zeta$ is independent of the position, we drop the position dependence of $s_i$ and write it as $s$.  Eq.(\ref{rw1}) is then rewritten as
\begin{equation}
 \frac{\Delta \pi_i(t)}{\Delta t} = -a_{i+1} \pi_{i+1}(t) + a_{i-1} \pi_{i-1}(t) + s \left( \pi_{i-1}(t ) +\pi_{i+1}(t ) -  2 \pi_{i}(t)\right). \label{rw2} 
\end{equation}
In the continuum limit $\ell \to 0$, $\pi_i (t)$ also shrinks to zero,  and it is no longer a meaningful quantity. Instead, we have to consider the probability density $\rho(x,t)$, which is related to $\pi_i(t)$ by
\begin{equation}
\rho(x) \equiv \frac{\pi_i}{\ell}, \quad x \equiv \ell i.
\end{equation}
By making identification 
\begin{equation}
F(x)= 2 \ell a_i,\quad B = s \ell^2, \label{id} 
\end{equation}
we recover the Fokker-Planck equation Eq.(\ref{fp}).
Note that $a_i$ and $s$ depend on $\Delta t$. Eq.(\ref{id}) shows that in order to obtain the correct limit, $\ell$ and $\Delta t$ should approach zero such that $a_i$ and $s$ diverge as 
\begin{equation}
a_i \sim \ell^{-1},\quad s \sim \ell^{-2}. \label{scaling}
\end{equation}

The Schnakenberg entropy production formula now becomes
\begin{eqnarray}
\Sigma &=&  \sum_{i}\sum_{j= i \pm 1} \pi_i(t) k_{i \to j} \log \left(\frac{ \pi_i(t) k_{i \to j}}{\pi_j(t)  k_{j \to i}} \right)\nonumber\\
&=&  \sum_{i}\sum_{j= i \pm 1} \pi_i(t) k_{i \to j} \log \left(\frac{ \pi_i(t) }{\pi_j(t) } \right)\nonumber\\
&+&   \sum_{i}  \pi_i(t) a_{i}\left[ \log \left(\frac{  k_{i \to (i+1)}}{  k_{(i+1) \to i}} \right) - \log \left(\frac{  k_{i \to (i-1)}}{ k_{(i-1) \to i}}  \right) \right]\nonumber\\
&+&  s \sum_{i}  \pi_i(t) \left[ \log \left(\frac{  k_{i \to (i+1)}}{  k_{(i+1) \to i}} \right) + \log \left(\frac{ k_{i \to (i-1)}}{  k_{(i-1) \to i}}  \right) \right]\nonumber\\
&=&  \sum_{i}\sum_{j= i \pm 1} \pi_i(t) k_{i \to j} \log \left(\frac{ \pi_i(t) }{\pi_j(t) } \right)\nonumber\\
&+&   \sum_{i}  \pi_i(t) a_{i}\left[ \log \left(\frac{ s + a_i}{s-a_{i+1}} \right) - \log \left(\frac{  s-a_{i}}{ s +a_{i-1}}  \right) \right]\nonumber\\
&+&  s \sum_{i}  \pi_i(t) \left[ \log \left(\frac{  s + a_i}{  s-a_{i+1}} \right) + \log \left(\frac{ s-a_{i}}{ s +a_{i-1}}  \right) \right]. \label{shk1}
\end{eqnarray}
 From Eq.(\ref{scaling}), we see that $a_i/s$ vanishes in the continuum limit as $a_i/s \sim \ell$, and consequently we can expand Eq.(\ref{shk1}) in powers of $a_i/s $. Since $a_i \sim \ell^{-1}$ and $s\sim \ell^{-2}$, the log term in the second and third lines of Eq.(\ref{shk1}) should be expanded up to the order of $O(\ell)$ and $O(\ell^2)$, respectively. Thus, we obtain
\begin{eqnarray}
\Sigma
&=&  \sum_{i}\sum_{j= i \pm 1} \pi_i(t) k_{i \to j} \left( \log  \pi_i(t) - \log  \pi_j(t)  \right)\nonumber\\
&+&   \sum_{i}  \pi_i(t) a_{i}\left[ 2 \frac{a_i}{s} +\frac{a_{i+1}}{s} + \frac{a_{i-1}}{s}+ O(\ell^2) \right]\nonumber\\
&+&   \sum_{i}  \pi_i(t) \left[ a_{i+1} - a_{i-1}  - \frac{a_{i}^2}{s}+\frac{a_{i+1}^2}{2 s}+\frac{a_{i-1}^2}{2 s} + O(\ell^3) \right]\nonumber\\
&\to& -\frac{d}{dt} \langle  \ln \rho(x,t) \rangle + \gamma \langle  \frac{F(x)^2}{T} \rangle + \langle \partial_x F(x) \rangle,
 \label{shk2}
\end{eqnarray}
where for simplicity we now consider the log function to be the natural log function. 
As in the case of the discrete state space, the heat production in the medium is~(appendix \ref{open})
\begin{equation}
q_{\rm tot} = T (\Sigma - \dot S_{\rm shan}) =  \gamma  \langle  {F(x)^2} \rangle + T \langle \partial_x F(x) \rangle
\end{equation}
where the Shannon entropy now takes the form\footnote{ $-\ln \rho(x,t)$ for  {\it a given position $x$} has been called the system entropy~\cite{seif1,seif2}. Similarly, $\gamma \dot x  F(x)/T$ for  {\it a given path} has been called the medium entropy production~\cite{seif1,seif2}. Therefore, we see that $S_{\rm shan}$ is the average of the system entropy over the ensemble of states, and $q_{\rm tot}/T = \Sigma - \dot S_{\rm shan}$ is the path ensemble average of the medium entropy production, by using Eq.(\ref{strat}).}
\begin{equation}
S_{\rm shan} \equiv -\langle \ln \rho(x,t) \rangle = -\int dx \rho(x,t) \ln \rho(x,t).
\end{equation}

As in the case of the discrete state space, the total heat production is decomposed into the house-keeping and excess heat production:
\begin{equation}
q_{\rm tot} = q_{\rm hk} + q_{\rm ex},
\end{equation}
where the excess heat production is obtained as in the case of the discrete state space~(Appendix \ref{open}):
\begin{equation}
q_{\rm ex} = T(S_{\rm cyc} -S_{\rm shan}) = T \frac{d}{dt}\langle \int dx \ln \rho^{\rm st}(x) \rangle, \label{qexc}
\end{equation}
with $\rho^{\rm st}(x)$ being the continuum limit of $\pi^{\rm st}_i$.
Consequently, we get
\begin{equation}
q_{\rm hk} =    \gamma  \langle  {F(x)^2} \rangle + T \langle \partial_x F(x) \rangle - T \frac{d}{dt}\langle \int dx \ln \rho^{\rm st}(x) \rangle. \label{qhkc}
\end{equation}

Here, $q_{\rm hk}$ and $q_{\rm ex}$, as given in Eq.(\ref{qhkc}) and Eq.(\ref{qexc}) respectively, are time-derivatives of the path ensemble averages of the housekeeping and excess heats defined by Hatano and Sasa~\cite{hs} for {\it a given path.} The details are as follows.  
The house-keeping and excess heat generated during a time-interval $[0,\tau]$ are defined in ref.\cite{hs} as
\begin{eqnarray}
Q_{\rm hk} &=& \int_0^\tau dt (f - \partial_x U - T \partial_x \ln \rho^{\rm st} (x; \lambda,f) ) \dot x \nonumber\\
Q_{\rm ex} &=&T \Delta \ln \rho^{\rm st} (x; \lambda,f) +  T \int_0^\tau dt \dot \lambda \partial_\lambda \ln \rho^{\rm st} (x; \lambda,f) + T \int_0^\tau dt \dot f \partial_f \ln \rho^{\rm st} (x; \lambda,f) \label{h1} 
\end{eqnarray}
where $\rho^{\rm st} (x; \lambda,f)$ is the steady state distribution of the Fokker-Planck equation obtained for fixed values of  $\lambda$ and $f$.  Because the focus of the current work is on the entropy production of a time-homogeneous Markov process, we assume $\lambda$ and $f$ to be time-independent. In this case, Eq.(\ref{h1}) is simplified to
\begin{eqnarray}
Q_{\rm hk} &=& \gamma \int_0^\tau dt \dot x F(x) - T \Delta \ln \rho^{\rm st} (x)  \nonumber\\
Q_{\rm ex} &=&T \Delta \ln \rho^{\rm st} (x) \label{h2} 
\end{eqnarray} 
where the dependence of $\rho^{\rm st}$ on $\lambda$ and $f$ is not explicitly shown, for the simplicity of notation. The path average of the $\dot x F(x)$ term in $Q_{\rm hk}$ is computed by the path integral using Stratonovitch discretization, leading to~\cite{seif1,seif2}
\begin{equation}
\langle F(x)\dot x \rangle = \langle F(x)^2  - B  F(x) {\rho(x,t)}^{-1} \partial_x \rho(x,t) \rangle,\label{strat}
\end{equation}
and we get
 \begin{eqnarray}
\langle Q_{\rm hk} \rangle &=& \gamma \langle \int_0^\tau dt  F(x)^2 \rangle +T  \langle \int_0^\tau dt  \partial_x F(x) \rangle - T \langle \Delta \ln \rho^{\rm st} (x) \rangle \nonumber\\
\langle Q_{\rm ex} \rangle &=&T \langle \Delta \ln \rho^{\rm st} (x) \rangle \label{h3} 
\end{eqnarray}  
where we have used the fact that
\begin{equation}
\langle   {\rho(x,t)}^{-1}{\partial_x \rho(x,t)} F(x) \rangle = \int dx \partial_x \rho(x,t) F(x) = -\int dx  \rho(x,t) \partial_x F(x) = -\langle  \partial_x F(x) \rangle
\end{equation}
 to derive the second term of the first line. By comparing Eq.(\ref{h3}) with Eq.(\ref{qexc}) and Eq.(\ref{qhkc}), we find that
 \begin{equation}
 q_{\rm hk} = \frac{d \langle Q_{\rm hk} \rangle}{d \tau} |_{\tau = t}, \quad q_{\rm ex} = \frac{d \langle Q_{\rm ex} \rangle}{d \tau} |_{\tau = t}. 
 \end{equation}

\section{The relation of the entropy components to nonadiabatic and adiabatic entropy productions~\cite{episto}}\label{adna}
A time-dependent Markov process defined by the transition rate $W^{\nu}_{i \to j} (\lambda_t)$, with $t \in [0,T]$, has been considered by Esposito and Van den Broeck\cite{episto}. Here, $\lambda_t$ is a time-dependent parameter that controls the dynamics of the system, which changes with time according to a fixed schedule. Multiple sources affecting the transition are considered, and the index $\nu$ denotes the mechanism responsible for the transition. A given path with $N$ jumps occurring at times $\tau_j, (j=1, \cdots N)$ due to mechanisms $\nu_j$ was considered, with the states of the system being $m_j$ for $\tau_{j-1} < t < \tau_{j}$, with $\tau_0 \equiv 0$ and  $\tau_{N+1} \equiv T$. For such a trajectory, the total entropy production was defined by 
\begin{equation}
\Delta S_{\rm tot}= \ln \frac{\pi_{m_0}(0)}{\pi_{m_N}  (T) }
 + \sum_{j=1}^N \ln \frac{ W^{\nu_j}_{m_{j-1} \to m_j}(\lambda_{\tau_j})}{W^{\nu_j}_{m_{j} \to m_{j-1}}(\lambda_{\tau_j})},\label{epitot}
 \end{equation}
which was then decomposed into the nonadiabatic entropy production $\Delta S_{\rm na}$ and the adiabatic entropy production $\Delta S_{\rm a}$: 
\begin{equation}
\Delta S_{\rm tot}= \Delta S_{\rm na} + \Delta S_{\rm a},
\end{equation}
with 
\begin{eqnarray}
\Delta S_{\rm na} &=&  \ln \frac{\pi_{m_0}(0)}{\pi_{m_N} (T)} 
+ \sum_{j=1}^N \ln \frac{\pi^{\rm st}_{m_j}(\lambda_{\tau_j}) }{\pi^{\rm st}_{m_{j-1}}(\lambda_{\tau_{j-1}})} \nonumber\\   
\Delta S_{\rm a} &=& \sum_{j=1}^N \ln  \frac{ W^{\nu_j}_{m_{j-1} \to m_j}(\lambda_{\tau_j}) \pi^{\rm st}_{m_{j-1}} (\lambda_{\tau_{j-1}}) }{W^{\nu_j}_{m_{j} \to m_{j-1}}(\lambda_{\tau_j}) \pi^{\rm st}_{m_j}(\lambda_{\tau_j})}. \label{sa}
\end{eqnarray}
Since the focus of the current article is on the cyclic time-homogeneous Markov model, we consider a special case where $\lambda$ is time-independent, and derive the relation between the entropy production components given by Eq.(\ref{sa}) and those given in this work. Because $\lambda$ is a fixed constant and plays no important role, the $\lambda$-dependence can be dropped from the notation. Furthermore, because we make no distinctions between various mechanisms causing the transition, but rather combine all of these effects into one transition rate, the $\nu$ dependence can be removed, and the transition rate can simply be written as $k_{i \to j}$. Under these assumptions, the expression in Eq.(\ref{sa}) is simplified to
\begin{eqnarray}
\Delta S_{\rm na} &=&  \ln \frac{\pi_{m_0}(0)}{\pi_{m_N} (T)} 
+ \ln \frac{\pi^{\rm st}_{m_N}} {\pi^{\rm st}_{m_0}}\nonumber\\
&=&  \ln \frac{\pi_{m_0}(0) \pi^{\rm st}_{m_N}}{\pi_{m_N} (T) \pi^{\rm st}_{m_0}} \nonumber\\
\nonumber\\   
\Delta S_{\rm a} &=& \sum_{j=1}^N \ln  \frac{ k_{m_{j-1} \to m_j}   }{ k_{m_{j} \to m_{j-1}} }+ \ln \frac{\pi^{\rm st}_{m_0}} {\pi^{\rm st}_{m_N}}\nonumber\\  
&=& \ln  \frac{\pi^{\rm st}_{m_0} \prod_{j=1}^N  k_{m_{j-1} \to m_j}   }{\pi^{\rm st}_{m_N} \prod_{j=1}^N k_{m_{j} \to m_{j-1}} }, \label{sa2}
\end{eqnarray}
where $\pi^{\rm st}_{m_j}$ with $1 \leq j \leq N-1$ from the denominator and numerator cancelled each other to obtain these expressions.
Note that the quantities in Eq.(\ref{sa2}) are defined for a given trajectory. Because entropy production for an ensemble of such trajectories is considered in the current work, we average $\Delta S_{\rm na}$ and $\Delta S_{\rm a}$ over path probabilities in order to make comparisons. We also shift the time interval from $[0,T]$ to $[t,t+T]$ without loss of generality and take the limit $T \to 0$. We then obtain 
\begin{eqnarray}
\langle \dot {S_{\rm na}} \rangle &\equiv& \lim_{T \to 0} \langle \Delta S_{\rm na} \rangle/T =  \lim_{T \to 0} \sum_{m,n}  \frac{\pi_m(t) P(m \to n; T)}{T}\ln \frac{\pi_m (t) \pi^{\rm st}_{n}}{\pi_{n} (t+T) \pi^{\rm st}_{m}},\nonumber\\    
\langle \dot {S_{\rm a}} \rangle &\equiv& \lim_{T \to 0} \langle \Delta S_{\rm a} \rangle/T \nonumber\\
&=&   \lim_{T \to 0} \frac{1}{T}  \sum_{N=0}^\infty \sum_{m}  \sum_{m_1 \ne m} \sum_{m_2 \ne m_1} \cdots \sum_{m_{N-1} \ne m_{N-2}} \sum_{n \ne m_{N-1}} \int_0^{T} d\tau_1 \int_{\tau_1}^{T} d\tau_2 \cdots \int_{\tau_{N-1}}^T d\tau_N \nonumber\\
&\times& \pi_m(t) P(m \to m_1; \tau_1) \cdots P(m_{N-1} \to n; \tau_N) \ln  \frac{\pi^{\rm st}_{m} k_{m \to m_1} \cdots k_{m_N \to n}   }{\pi^{\rm st}_{n} k_{n \to m_{N-1}} \cdots k_{m_1 \to m} }. \label{sa3}
\end{eqnarray}
Here, $P(m \to n; T)= (\exp K T)_{m n}$ is the conditional probability that the system is at the state $n$ at time $t+T$, given that it is at the state $m$ at time $t$, where $K$ is the matrix whose $(m,n)$ component is $k_{m \to n}$. For small $T$, we have the expansion
\begin{equation}
(\exp K T)_{m n} = \delta_{m n} + T k_{m n} + O(T^2).
\end{equation}
Therefore, we get
\begin{eqnarray}
\langle \dot {S_{\rm na}} \rangle &=&   \sum_{m,n} \pi_m(t) k_{m \to n} \ln \frac{\pi_m (t) \pi^{\rm st}_{n}}{\pi_{n} (t) \pi^{\rm st}_{m}}+ \lim_{T \to 0} \frac{1}{T} \sum_{n} \pi_n(t) (\ln \pi_n (t) - \ln \pi_{n} (t+T) ) \nonumber\\  
&=& \sum_{m,n} \pi_m(t) k_{m \to n} \ln \frac{\pi_m (t) \pi^{\rm st}_{n}}{\pi_{n} (t) \pi^{\rm st}_{m}}- \sum_{n}\dot \pi_n (t)\nonumber\\
&=& \sum_{m,n} \pi_m(t) k_{m \to n} \ln \frac{\pi_m (t) \pi^{\rm st}_{n}}{\pi_{n} (t) \pi^{\rm st}_{m}} ,  \nonumber\\
\langle \dot {S_{\rm a}} \rangle &\equiv& \lim_{T \to 0} \langle \Delta S_{\rm a} \rangle/T =   \lim_{T \to 0}   \sum_{N=0}^\infty \frac{T^{N-1}}{N!} \sum_{m} \sum_{m_1 \ne m} \sum_{m_2 \ne m_1} \cdots \sum_{m_N \ne m_{N-1}} \sum_{n \ne m_{N}}  \nonumber\\
&\times& \pi_m(t) k_{m \to m_1} \cdots k_{m_{N-1} \to n} \ln  \frac{\pi^{\rm st}_{m}   k_{m \to m_1} \cdots k_{m_N \to n} }{\pi^{\rm st}_{n} k_{n \to m_{N-1}} \cdots k_{m_1 \to m} } \nonumber\\
 &=&  \sum_{m \ne n} \pi_m(t) k_{m \to n} \ln  \frac{ \pi^{\rm st}_{m} k_{m \to n} }{\pi^{\rm st}_{n} k_{n \to m} }\nonumber\\
 &=&  \sum_{m, n} \pi_m(t) k_{m \to n} \ln  \frac{ \pi^{\rm st}_{m}  k_{m \to n}  }{\pi^{\rm st}_{n} k_{n \to m} }. \label{sa4}
\end{eqnarray} 
 
Comparing these expressions with
\begin{eqnarray}
\dot S_{\rm cyc} &=& -\sum_i {\dot \pi}_i(t) \left( \ln \left(\frac{\pi_i(t)}{\pi^{\rm st}_i}\right) + 1 \right)\nonumber\\
&=& -\sum_{i,j}(\pi_j(t)k_{j \to i} - \pi_i(t)k_{i \to j}) \left( \ln \left(\frac{\pi_i(t)}{\pi^{\rm st}_i}\right) + 1 \right)\nonumber\\
&=& \sum_{i,j} \pi_i(t)k_{i \to j} \ln \left(\frac{\pi_i(t) \pi^{\rm st}_j}{\pi_j(t) \pi^{\rm st}_i}\right)\label{dscy}
\end{eqnarray}
and
\begin{eqnarray}
\dot S_{\rm hk} &=& \Sigma - \dot S_{\rm cyc}  \nonumber\\
&=& \sum_{i,j} \pi_i(t) k_{i \to j} \log \left(\frac{ \pi_i(t) k_{i \to j}}{\pi_j(t)  k_{j \to i}} \right) - \sum_{i,j} \pi_i(t)k_{i \to j} \ln \left(\frac{\pi_i(t) \pi^{\rm st}_j}{\pi_j(t) \pi^{\rm st}_i}\right)\nonumber\\
&=& \sum_{i,j} \pi_i(t) k_{i \to j} \log \left(\frac{ \pi^{\rm st}_i k_{i \to j}}{\pi^{\rm st}_j  k_{j \to i}} \right), \label{dshk}
\end{eqnarray}
in the cyclic regime, we immediately see that $\langle \dot {S_{\rm na}} \rangle = \dot S_{\rm cyc}$ and $\langle \dot {S_{\rm a}} \rangle = \dot S_{\rm hk} $. Note that because the driver degrees of freedom are not explicitly incorporated in the model, $S_{\rm tot}$ in Eq.(\ref{epitot}) does not include the contribution from the hidden entropy.

\newpage
\begin{figure}
\includegraphics[width=\columnwidth]{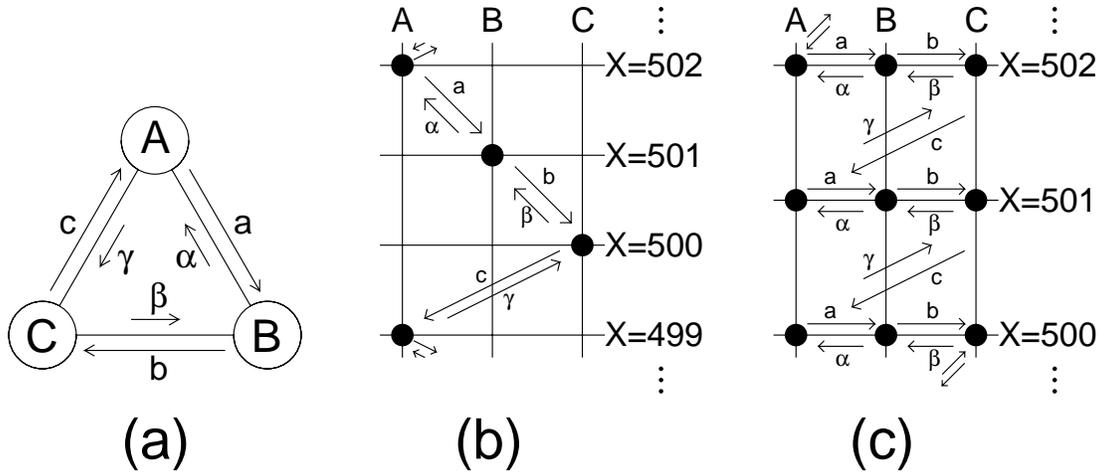}
\caption{(a) Cyclic Markov model with three states. (b),(c) Examples of extended models where an additional degree of freedom $X$ is introduced. }
\label{model} 
\end{figure}
\begin{figure}
\includegraphics[height=18cm]{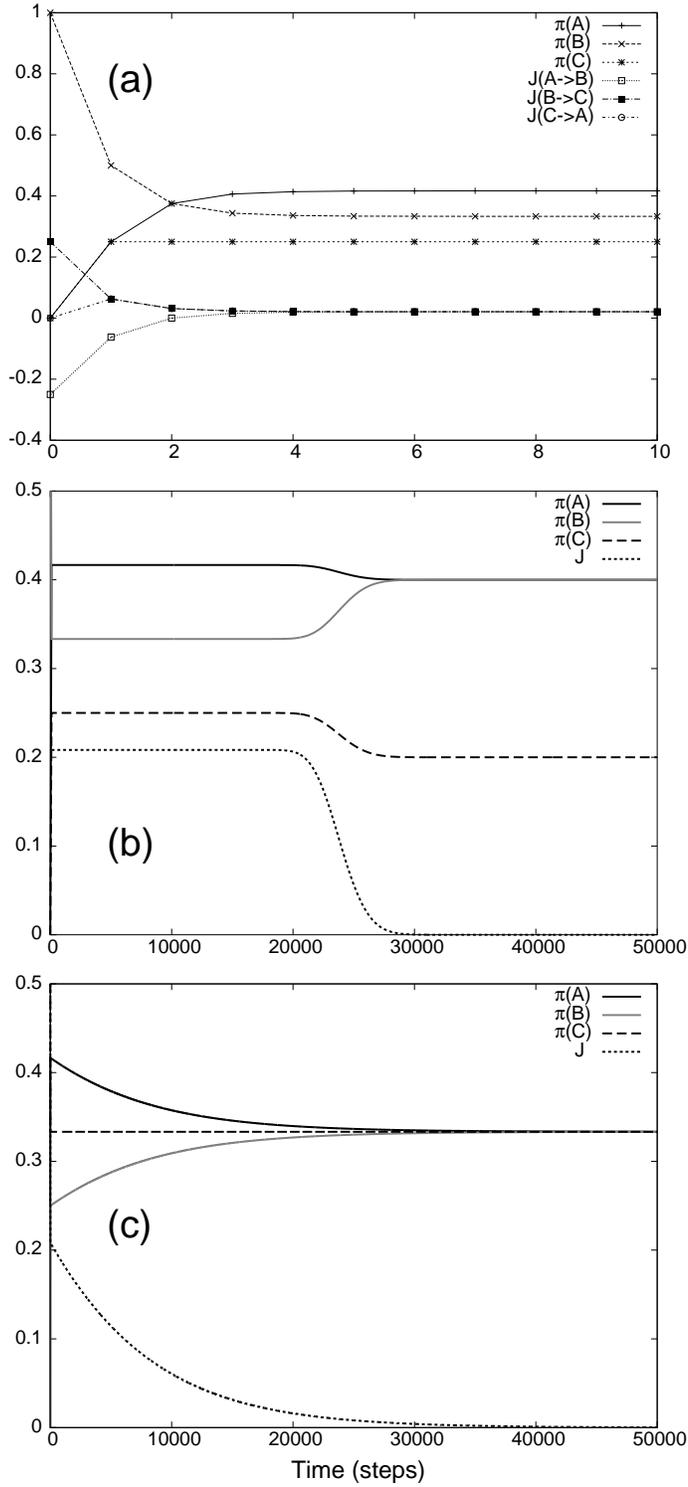}
\caption{Time evolution of three-state model with discrete time. Model 1: $a=b=\alpha=\beta=\gamma=0.25/\Delta t$, and $c =0.5/\Delta t$. Model 2: $a=b=\alpha=\beta=0.25/\Delta t$, $c =1.0 (X/N)\Delta t^{-1}$, and $\gamma=0.5(1-X/N)\Delta t^{-1}$. The probability distribution and currents are displayed as functions of time. (a)  Probability distribution and currents of both models at early times. (b)  Probability distribution and currents of model 1 at late times. For better visibility, $J$ was multiplied by 10. (c) Probability distribution and currents of model 1 at late times. For better visibility, $J$ was multiplied by 10. }
\label{steady} 
\end{figure}
\begin{figure}
\includegraphics[height=18cm]{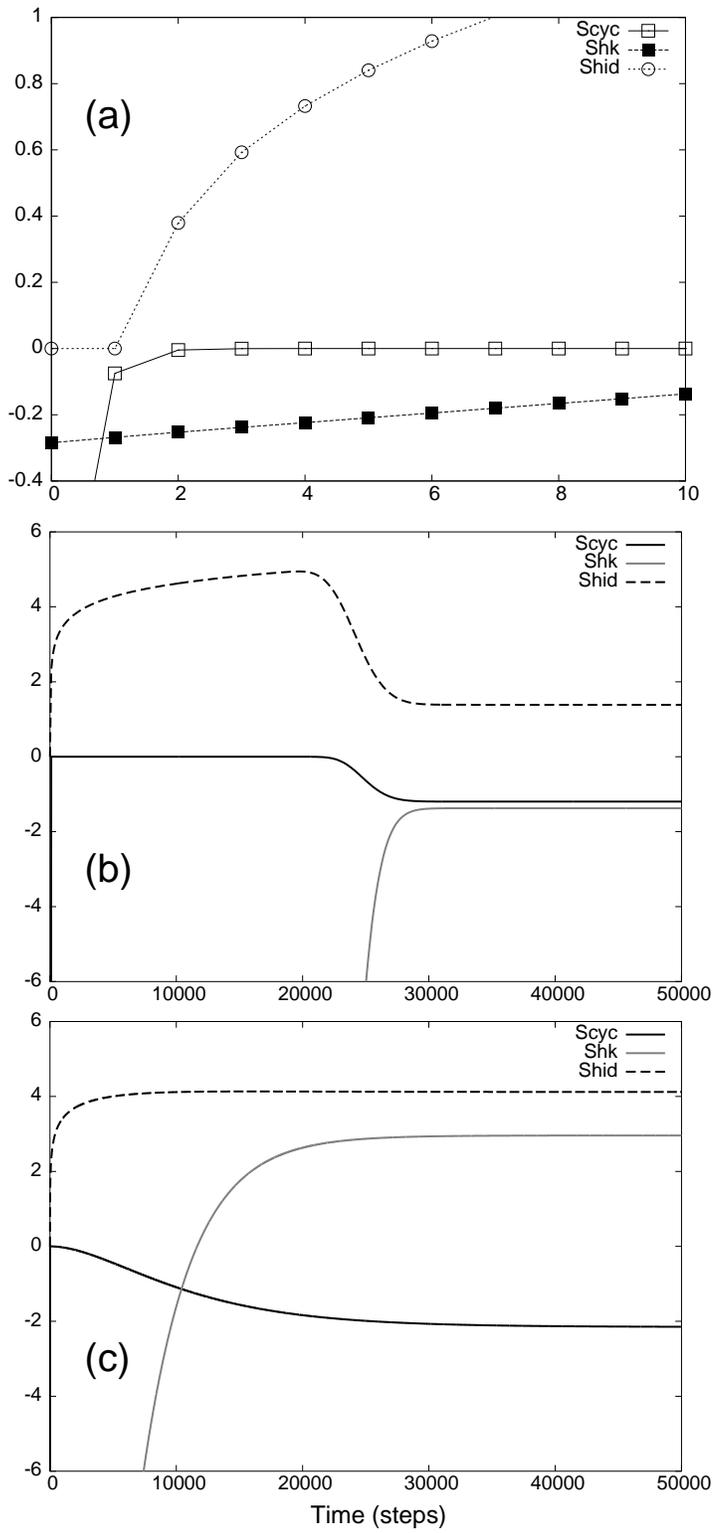}
\caption{Time evolution of three-state model with discrete time. Model 1: $a=b=\alpha=\beta=\gamma=0.25/\Delta t$, and $c =0.5/\Delta t$. Model2: $a=b=\alpha=\beta=0.25/\Delta t$, $c =1.0 (X/N)\Delta t^{-1}$, and $\gamma=0.5(1-X/N)\Delta t^{-1}$. The entropy components are displayed as functions of time. (a) Entropy components of both models at early times. Separate constants were added to $S_{\rm hk}$ of the two models, so that their graphs are superposed. (b) Entropy components of model 1 at late times. (c) Entropy components of model 2 at late times. For each model, a constant was added to $S_{\rm hk}$ for easier comparison, and $S_{\rm cyc}$  was multiplied by 100 for better visibility.}
\label{entropy} 
\end{figure}
\begin{figure}
\includegraphics[width=20cm]{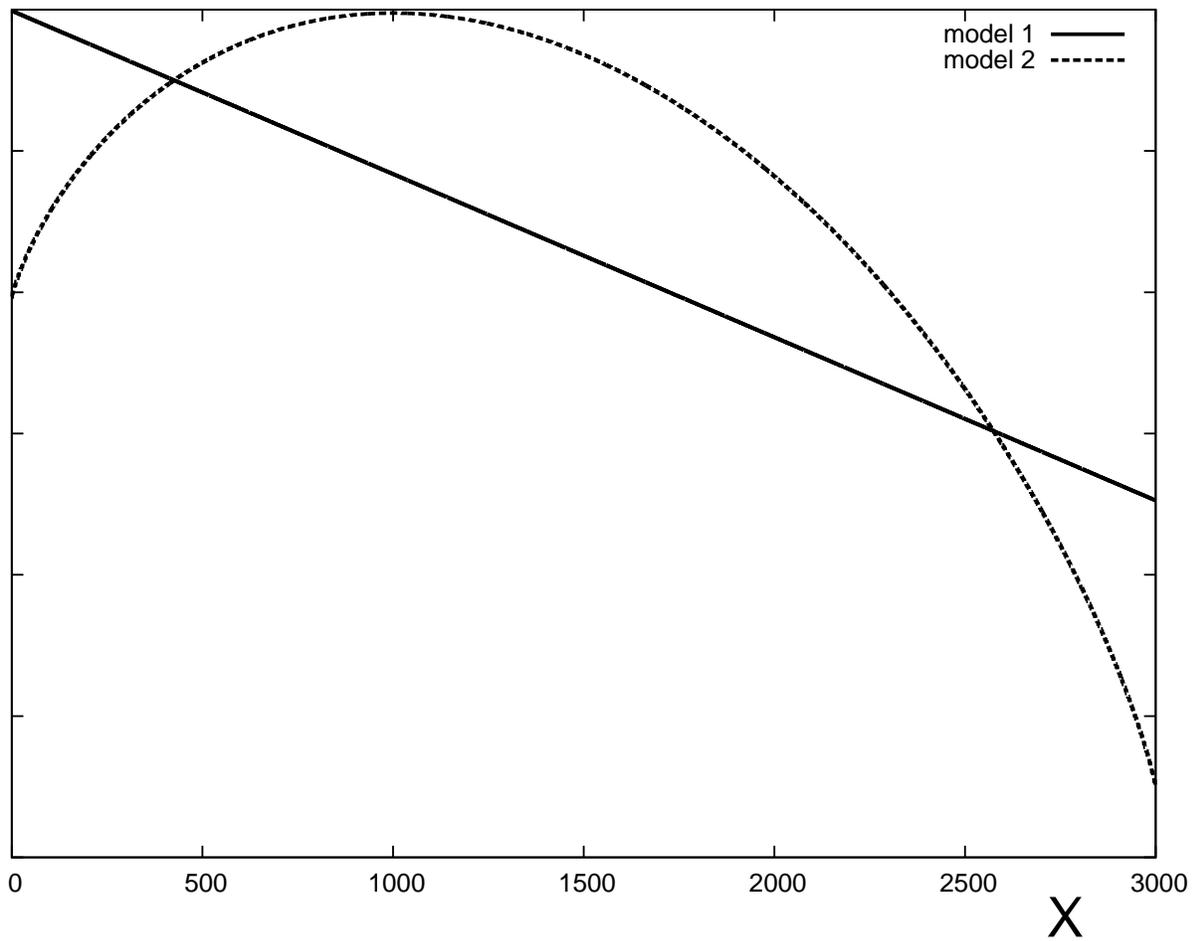}
\caption{Equilibrium distribution of model 1 and model 2 as a function of $X$, shown in log scale. Parameters are set to the values used for generating Figures \ref{steady} and \ref{entropy}.}
\label{equi} 
\end{figure}
\end{document}